\documentclass[reprint,superscriptaddress,amsmath,amssymb,aps,floatfix]{revtex4-1}
 
\usepackage{graphicx}
\usepackage{dcolumn}
\usepackage{bm}
\usepackage{url} 

\begin{document}


\title[NESSIAS Effect at Semiconductor Nanostructures]{Origin and Quantitative Description of the NESSIAS Effect at Si Nanostructures} 

\author{Dirk K{\"o}nig}
\affiliation{Integrated Materials Design Lab (IMDL), The Australian National University, ACT 2601 Canberra, Australia}
\altaffiliation{Smart Materials and Surface Group, University of New South Wales, NSW 2052 Sydney, Australia}
\email{solidstatedirk@gmail.com}
\author{Michael Frentzen}
\affiliation{Institute of Semiconductor Electronics (IHT), RWTH Aachen University, 52074 Aachen, Germany}
\author{Daniel Hiller}
\affiliation{Institute of Applied Physics (IAP), Technische Universit\"{a}t Bergakademie Freiberg, 09599 Freiberg, Germany}
\author{No{\"e}l Wilck}
\affiliation{Institute of Semiconductor Electronics (IHT), RWTH Aachen University, 52074 Aachen, Germany}
\author{Giovanni Di Santo}
\author{Luca Petaccia}
\affiliation{Elettra Sincrotrone Trieste, Strada Statale 14 km 163.5, 34149 Trieste, Italy}
\author{Igor P\'{i}\v{s}}
\author{Federica Bondino}
\author{Elena Magnano}
\affiliation{IOM-CNR, Instituto Officina dei Materiali, Area Science Park S.S. 14 km 163.5, 34149 Trieste, Italy}
\altaffiliation[E. Magnano:\ ]{Department of Physics, University of Johannesburg, PO Box 524, Auckland Park 2006, South Africa}
\author{Joachim Mayer}
\affiliation{Ernst-Ruska Centre for Microscopy and Spectroscopy with Electrons, RWTH Aachen University, 52074, Germany}
\author{Joachim Knoch}
\affiliation{Institute of Semiconductor Electronics (IHT), RWTH Aachen University, 52074 Aachen, Germany}
\author{Sean C. Smith}
\affiliation{Department of Applied Mathematics, Research School of Physics and Engineering, The Australian National University, ACT 2601 Canberra, Australia}
\altaffiliation{Integrated Materials Design Lab (IMDL), The Australian National University, ACT 2601 Canberra, Australia}

\begin{abstract}
The electronic structure of low nanoscale (LNS) intrinsic silicon (i-Si) embedded in SiO$_2$ \emph{vs.} Si$_3$N$_4$ shifts away from \emph{vs.} towards the vacuum level $E_{\rm vac}$, as described by the Nanoscale Electronic Structure Shift Induced by Anions at Surfaces (NESSIAS). Here, we fully explain the NESSIAS based on the quantum chemical properties of the elements involved. Deriving an analytic parameter $\Lambda$ to predict the highest occupied molecular orbital energy of Si nanocrystals (NCs), we use various hybrid-DFT methods and NC sizes to verify the accuracy of $\Lambda$. 
We report on first experimental data of Si nanowells (NWells) embedded in SiO$_2$ \emph{vs.} Si$_3$N$_4$ by X-ray absorption spectroscopy in total fluorescence yield mode (XAS-TFY) which are complemented by ultraviolet photoelectron spectroscopy (UPS), characterizing their conduction band and valence band edge energies $E_{\rm C}$ and $E_{\rm V}$, respectively. Scanning the valence band sub-structure by UPS over NWell thickness, we derive an accurate estimate of $E_{\rm V}$ shifted purely by spatial confinement, and thus the actual $E_{\rm V}$ shift due to NESSIAS. For 1.9 nm thick NWells in SiO$_2$ \emph{vs.} Si$_3$N$_4$, we get offsets of $\Delta E_{\rm C}=0.56$ eV and $\Delta E_{\rm V}=0.89$ eV, demonstrating a type II homojunction in LNS i-Si. This p/n junction generated by the NESSIAS eliminates any deteriorating impact of impurity dopants, offering undoped ultrasmall Si electronic devices with much reduced physical gate lengths and CMOS-compatible materials.
\end{abstract}

\maketitle
\section{\label{intro}Introduction}
Silicon (Si) nanowells (NWells) with a thickness of $d_{\rm Well}\leq\ {\rm ca.}\ 3.3$ nm embedded in silicon dioxide (SiO$_2$) \emph{vs.} silicon nitride (Si$_3$N$_4$) show an electronic structure shift with respect to the vacuum energy level $E_{\rm vac}$ as measured by ultraviolet photoelectron spectroscopy (UPS) and  X-ray absorption spectroscopy in total fluorescence yield mode (XAS-TFY) \cite{Koe18a,Koe19,Koe21}. NWells embedded in SiO$_2$ (Si$_3$N$_4$) get shifted to higher (lower) binding energies, that is, away from (towards to) $E_{\rm vac}$. This Nanoscopic Electronic Structure Shift Induced by Anions at Surfaces (NESSIAS) effect is caused by quantum chemical properties of the anions forming the dielectric which surrounds the low nanoscale (LNS) Si. While the NESSIAS effect has been established in theory and experiment, its exact origin and quantitative description are still elusive. Here, we deliver a detailed quantum chemical explanation of the NESSIAS effect, complemented with its semi-quantitative description which serves to predict NESSIAS in LNS intrinsic Si (i-Si) for a variety of anions in embedding/coating dielectrics. To this end, we provide experimental evidence and details of the quantum-chemical concept which leads to the NESSIAS effect.

The NESSIAS effect induces a p/n junction on semiconductor nanostructures such as fins, nanowires (NWires), or nanocrystals (NCs) by enabling an electron flooding of the nanostructure when coated with SiO$_2$ \cite{Koe21}, or a virtually complete electron drainage from the nanostructure when coated with Si$_3$N$_4$ \cite{Koe18a,Koe19}, introducing a high density of holes into the nanovolume by the latter process. This re-arrangement of charge carrier densities has far-reaching consequences for semiconductor devices in very large scale integration (VLSI), ultra-low power and cryo-electronics. Spatial fluctuations of dopant densities, out-diffusion and self-purification impose a size limit onto VLSI devices as evident from \emph{physical} gate lengths hovering around 20 nm since ca. 2014 \footnote{It is interesting to note in this context that the VLSI technology nodes do not reflect the physical gate length. We have a planar MOSFET \emph{model} shrunken to a size where it \emph{would perform} as the fin-FET of the respective technology node \cite{Colli16}.}. With thermal dopant ionization not required, junctions induced by the NESSIAS effect should remain fully functional down to extremely low temperatures as useful for peripheral electronics in qbit manipulation \cite{Ladd10}.

The detection of the NESSIAS in LNS i-Si requires an absolute assignment of energies to $E_{\rm vac}$, combined with elaborate UPS and XAS-TFY measurements and refined data processing for improved signal-to-noise ratios \cite{Koe18a,Koe19,Koe21}. To this end, the structures under investigation have to be in the range of the NESSIAS impact length. Many published DFT calculations \cite{Cast15,Jaro17,Ossi20,Hali20} lack an energy assignment on an absolute scale. Possibly identical LNS i-Si NWell systems near the low end of the one-digit nm range embedded in different dielectrics pose a challenge in experiment. The common perception of Si$_3$N$_4$ as an inferior dielectric on grounds of interface defect density \cite{Koe22} and its more complex technology as opposed to SiO$_2$ \cite{Lee12} are likely reasons for the literature on LNS Si embedded in or coated with Si$_3$N$_4$ being rather scarce. Indeed, standard Si$_3$N$_4$ has an interface defect density to LNS Si which exceeds values of SiO$_2$/Si interfaces \emph{ca.} 13-fold \cite{Thoa11,Basa00,Koe22}, though refined preparation techniques for high-quality H-passivated Si$_3$N$_4$-coatings rival trap densities on SiO$_2$/Si interfaces \cite{Jung11}. This complex situation may explain why the NESSIAS might have been overlooked in the past. 

After introducing the methods used in Section \ref{meth}, we deliver a phenomenological and qualitative explanation of the NESSIAS effect in Section \ref{qualiExplain}, resorting to quantum chemical properties of involved chemical elements. In Section \ref{analytics}, we derive a semi-quantitative analytic parameter $\Lambda$ of the NESSIAS effect to describe the energy of the highest occupied molecular orbital (HOMO) $E_{\rm HOMO}$ as a function of anion-specific quantum chemical properties combined with the charge of these main anions $q_{\rm main}$ of the ligand groups attached to Si NCs. In Section \ref{DFT}, we test $\Lambda$ with a variety of density functionals (DFs) and most anionic elements of the first and second row of the periodic table with respect to Si, subject to the availability of experimental quantum chemical data.  
Our discussion is complemented with synchrotron data in Section \ref{SynchroExpr}. Since the electronic structure of NWells shifts as a function of quantum confinement (QC), it is essential to separate this phenomenon induced by a \emph{spatial} limit from the NESSIAS brought about by the quantum-chemical nature of embedding dielectric \emph{vs.} Si. To this end, we evaluate our results from synchrotron UPS measurements in Section \ref{QC-SubEdges}, revealing sub-structures of the valence band which serve to estimate the actual QC. This true QC is revealed by the shift of the valence band (VB) edge to higher binding energies as a sole function of NWell thickness, occurring in all samples irrespective of the embedding dielectric. We use this VB edge as a reference level to estimate the \emph{actual} NESSIAS as per embedding dielectric in the QC regime. 
Next, we evaluate our experimental data of the VB and conduction band (CB) edges as measured by synchrotron XAS-TFY in Section \ref{QC-FullPic}, establishing the link between bulk Si and SiO$_2$ \emph{vs.} Si$_3$N$_4$ as given by ultrathin Si NWells coated with the respective dielectric. Section \ref{WrapUp} delivers a conclusion. 

\section{\label{meth}Methods}
\subsection{\label{Prep-Meth}Sample Preparation}
After determining the Deal-Gove parameters \cite{Deal65} for the furnace oxidation of silicon-on-insulator (SOI) samples, the SOI crystalline Si (c-Si) layers (p-type, 1 $\Omega$cm) on Si wafers with 145 nm buried SiO$_2$ (BOx) were oxidized down to a thickness of 2.1 to 6.0 nm. The SiO$_2$ was removed by etching in a buffered oxide etch (BOE; 1 wt-\% HF buffered with NH$_4$F), followed by a self-limiting oxidation in 68 wt-\% HNO$_3$ at 120 $^{\circ}$C, yielding a 1.1 to 5.0 nm Si-NWell with 1.4 nm SiO$_2$ capping. A lateral metal contact frame was processed on the front surface by photolithographical structuring, wet-chemical etching in BOE for opening the top SiO$_2$ layer and thermal evaporation of 300 nm Al, followed by a lift-off in acetone. The Si reference samples were contacted directly on their front surface. NWell samples were coated with photo resist immediatley after NWell thickness measurements using Mueller matrix ellipsometry straight after processing to prevent oxidation in air. Resist was removed just prior to sample mounting at the beamline. Si reference samples consisted of (001)-Si wafer (Sb-doped $n$-type, 0.01 $\Omega$cm) which were treated with a BOE immediately before sample mounting under a N$_2$ gas flow with swift loading into the ultra-high vacuum (UHV) annealing chamber. 

Si-NWells in Si$_3$N$_4$ were processed in analogy to the ones in SiO$_2$, using the same SOI wafers as starting point.
The SOI was thinned down to a remanent Si device layer which was 1 nm thicker than the final NWell thickness, accommodating for Si consumption during the growth of Si$_3$N$_4$. Next, the SiO$_2$ capping is removed with BOE immediately before growing 3 nm of Si$_3$N$_4$ in an ammonia atmosphere by rapid thermal nitridation (RTN). Afterwards, 20 nm of Si$_3$N$_4$ and 80 nm of SiO$_2$ were deposited by plasma enhanced chemical vapor deposition, followed by chemical mechanical polishing of the SiO$_2$ layer. After an RCA clean \cite{Kern78}, the samples were bonded to a Si sample \cite{Kraeu98} covered by SiO$_2$ of ca. 1.5 nm thickness which was grown during RCA-SC2 step. The original Si substrate of the SOI samples was etched back using a cyclic deep reactive ion etching process based on passivation with C$_4$F$_8$ and etching with SF$_6$. The BOx served as a stopping layer and was subsequently removed
by a BOE immediately before growing 1 nm of Si$_3$N$_4$ by RTN in ammonia atmosphere. Contacts to the NWells were fabricated by photolithography, etching the Si$_3$N$_4$ with 1 wt-\% hydrofluroric acid and deposition of 300 nm Al.

\begin{figure}[h!]
\begin{center}
\includegraphics[width=8.6cm,keepaspectratio]{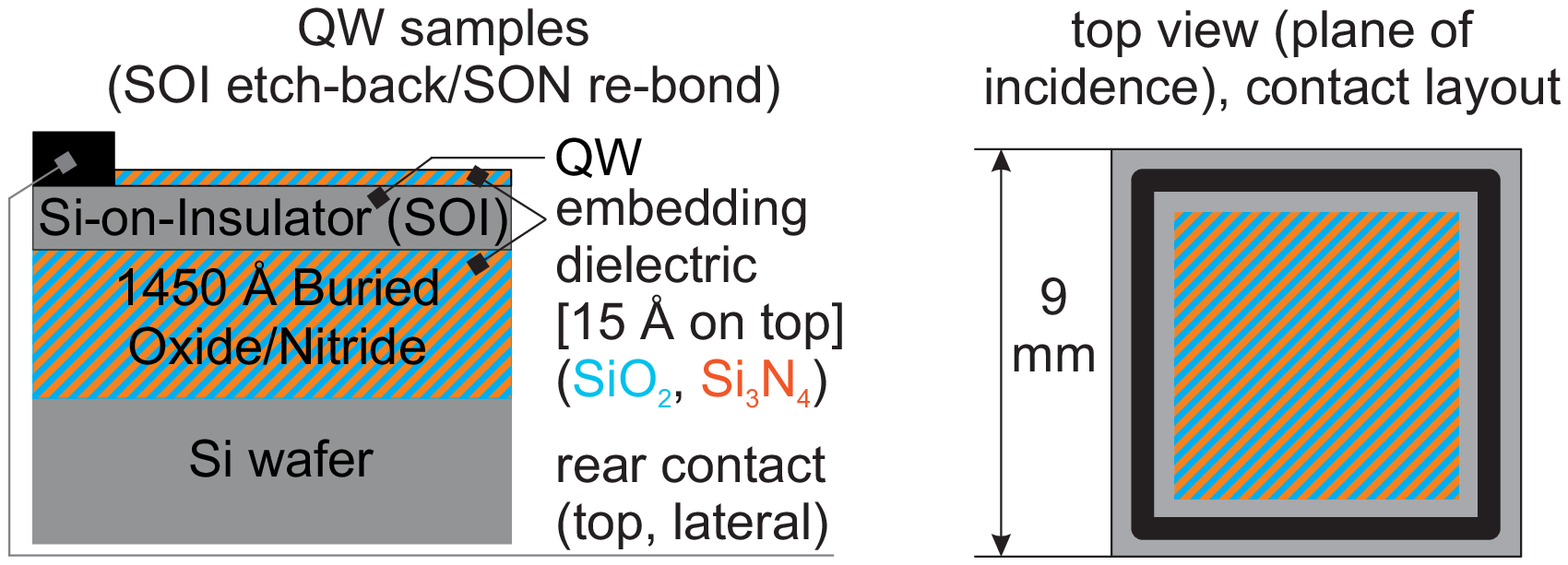}\\[0.1cm]
{\bf (a)}\\[0.3cm]
\includegraphics[width=8.6cm,keepaspectratio]{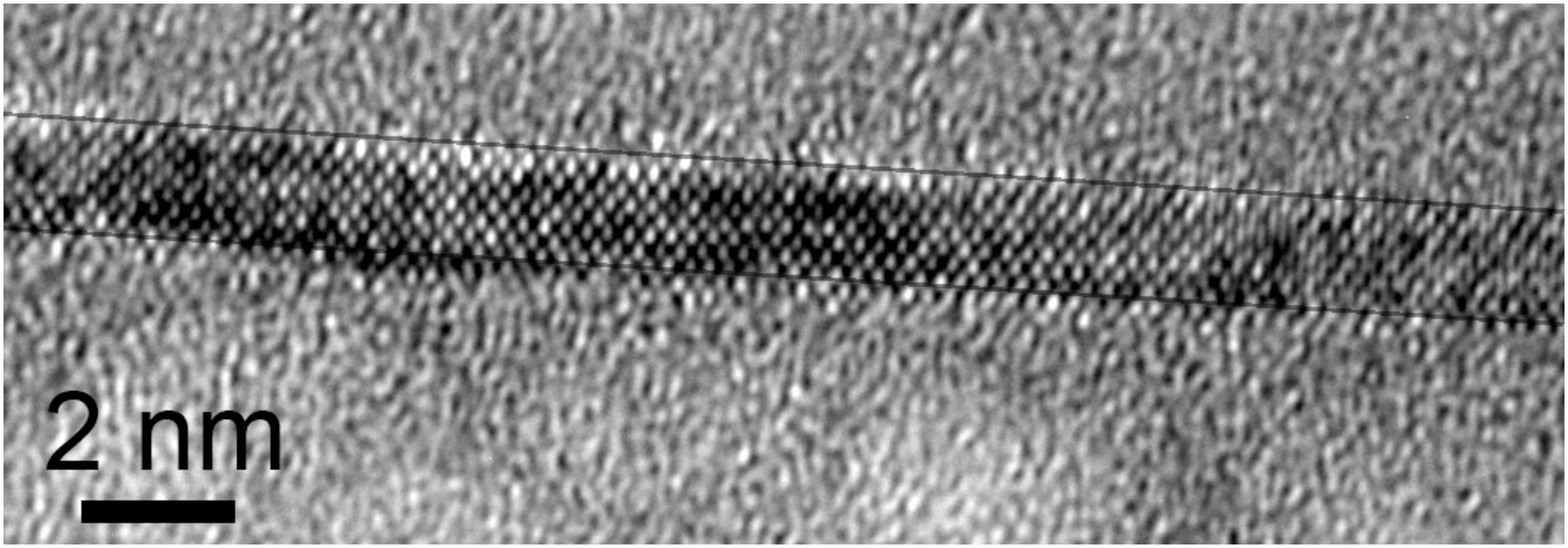}\\
{\bf (b)}
\vspace*{-0.188cm}
\end{center}
\caption{\label{fig01} Layout of Si$_3$N$_4$- and SiO$_2$-embedded Si NWell samples for synchrotron UPS and XAS-TFY, shown in X-section and top view (a). X-section view of 1.5 nm (001)-Si NWell in Si$_3$N$_4$ as obtained by TEM (b). Semi-transparent black lines show interfaces to adjacent Si$_3$N$_4$ from where the NWell thickness was determined.} 
\end{figure}

The layout of the samples and a high resolution TEM image of a NWell sample is shown Fig. \ref{fig01}.

\subsection{\label{Char-Meth} Characterization}
For synchrotron-based characterization techniques such as X-ray absorption spectroscopy measurements in total fluorescence yield (XAS-TFY) and UV photoelectron spectroscopy (UPS), all samples were subject to a UHV anneal for 60 min at 500 K to desorb water and air-related species from the sample surface prior to measurements.
Synchrotron XAS-TFY measurements were carried out at the BACH CNR beamline \cite{Zang04,Steb11} at the Elettra synchrotron radiation facility in Trieste, Italy, in top-up mode at an electron energy of 2.4 GeV (140 mA electron ring current). The photon energy scale has been calibrated using the energy position of the Si L3-edge (99.6 eV) measured on a reference (001)-Si wafer. A multitude of spectra was recorded per sample to obtain the data ensemble required for statistical post-processing.  Further information and data on measurement parameters, measurement geometry, and statistical data post-processing can be found in Ref. \cite{SuppInfo}. 

Synchrotron UPS measurements were realized at the BaDElPh beamline \cite{Peta09} at the Elettra synchrotron radiation facility in Trieste, Italy, in top-up mode at an electron energy of 2.0 GeV (300 mA electron ring current). Single scans of spectra were recorded over 12 h per NWell sample and subsequently added up for eliminating white noise.
Scans for the Si-reference sample were recorded over 2 h and subsequently added up. All NWell samples were excited with photon energies of initially 8.9 eV and of 8.0 eV for subsequent measurements, and a photon flux of ca. $2\times 10^{12}$ s$^{-1}$. The incident angle of the UV beam onto the sample was 50$^{\circ}$ with respect to the sample surface normal, the excited electrons were collected with an electron analyzer along the normal vector of the sample surface. Energy calibration of the UPS was realized using a tantalum (Ta) stripe in electrical contact to the sample as work function reference. For details on UPS data and measurement, we refer to the supporting information of \cite{Koe18a} and the appendix of \cite{Koe19}.

The thickness of Si-NWells and its adjacent SiO$_2$ layers  were measured using an ACCURION nanofilm ep4se ellipsometer \cite{Accurion2020} using a Mueller matrix approach \cite{Muel48,Azza78,Save09}.
Additional thickness measurements of the Si NWells and top SiO$_2$ were carried out by Transmission Electron Microscopy (TEM) on selected samples to confirm and/or calibrate ellipsometry measurements, see to representative NWell image in Fig. \ref{fig01}(b). All TEM samples were capped with a protective 100 nm thick SiO$_2$-layer to facilitate the preparation of X-sections by the focused ion beam (FIB) technique using a FEI Strata FIB 205 workstation. Some samples were further thinned by 
a Fischione NanoMill. TEM analysis of the X-sections was performed at a FEI Tecnai F20 TEM operated at 200 kV at the Central Facility for Electron Microscopy, RWTH Aachen University, and at the spherical aberration corrected FEI Titan 80-300 TEM operated at 300 kV at Ernst Ruska-Centre, Forschungszentrum J\"ulich \cite{TITAN}.

\subsection{\label{DFT-Meth} Density Functional Theory (DFT) Calculations}
Real space calculations were carried out with a molecular orbital basis set (MO-BS) and Hartree-Fock (HF)/DFT methods, employing the {\sc Gaussian09} program package \cite{G09} with the GaussView program \cite{GV5} for visualization. Initially, the MO-BS wavefunction ensemble was tested and optimized for describing the energy minimum of the system (variational principle; stable $=$ opt) with the HF method \cite{Hart28a,Hart28b,Fock30}. 
Exact exchange interaction inherent to HF is crucial in obtaining accurate bond geometries, see supporting information of \cite{Koe18a}. 
As MO-BS, we used the Gaussian type 3-21G MO-BS \cite{Gor82}. This HF/3-21G route was used for the structural optimization of approximants to obtain their most stable configuration (maximum integral over all bond energies); root mean square (RMS) and peak force convergence limits were 
15.3 meV\,{\AA}$^{-1}$ and 23.1 meV\,{\AA}$^{-1}$, (300 and 450 $\mu$Ha/$a_{\rm B,0}$), respectively. Optimized geometries were used to calculate their electronic structure by testing and optimizing the MO-BS wavefunction ensemble with the non-local hybrid DF B3LYP \cite{Beck88,Lee88,Beck93}, its modified form featuring the Coulomb Attenuation Method (CAM-B3LYP) for more accurate asymptotic non-local exchange interactions \cite{Yana04}, and the HSE06 hybrid DF with its parameters from 2006 \cite{Kruk06}.
As MO-BS, we used the Gaussian type 6-31G(d) MO-BS which contains d-polarization functions (B3LYP/6-31G(d)\,) \cite{Fra82} for all chemical elements. For all calculations, tight convergence criteria were set to the self-consistent field routine and no symmetry constraints to MOs were applied. Ultrafine integration grids were used throughout. The supporting information of \cite{Koe14,Koe18a} contain detailed accuracy assessments.

\section{\label{results}Results}
\subsection{\label{qualiExplain}Qualitative Explanation of the NESSIAS Effect}
When a common boundary between two different solids with a bandgap is formed, an interface charge transfer (ICT) occurs \cite{Moench01}, generating an interface dipole which may shift the electron work function of both materials with respect to the intrinsic solid-vacuum interface \cite{Moench01,Camp96}. The solid which accumulates extrinsic electrons experiences an electronic structure shift to lower binding energies $E_{\rm bind}$, hence to $E_{\rm vac}$, the other solid which provides the electronic charge experiences an electronic structure shift in the opposite direction. When LNS i-Si is coated with 1 ML SiO$_2$ or Si$_3$N$_4$, the ICT provides about the same amount of electrons to O and N as main interface anions in OH and NH$_2$ groups \cite{Koe19}, see to top section in Table \ref{tab1}.  
\begin{table}[h!]
  \caption{\mdseries{Electronic structure data of Si$_x$ NCs terminated with OH or NH$_2$ ($\equiv 1$ ML SiO$_2$, Si$_3$N$_4$) or 1.5 ML SiO$_2$ or Si$_3$N$_4$, see \cite{Koe19} for details, showing the NC size $d_{\rm NC}$, cumulative charge transferred from the Si NC into the dielectric $\sum q_{\rm \,ICT}$, and HOMO and LUMO energies $E_{\rm HOMO}$, $E_{\rm LUMO}$.}} 
  \label{tab1}
  \begin{tabular}{l|rrrr}
    approximant&$d_{\rm NC}$&$\sum q_{\rm\, ICT}$&$E_{\rm{HOMO}}$&$E_{\rm{LUMO}}$\\
    &[nm]&[e]&[eV]&[eV]\\
    \hline
    Si$_{10}$(NH$_2)_{16}$&&$-4.20$&$-4.02$&$-0.50$ \\ 
    Si$_{10}$(OH)$_{16}$&\raisebox{1.5ex}[-1.5ex]{0.72}&$-4.51$&$-5.31$&$-2.08$\\[0.1cm]
    Si$_{35}$(NH$_2)_{36}$&&$-10.10$&$-3.85$&$-0.66$ \\ 
    Si$_{35}$(OH)$_{36}$&\raisebox{1.5ex}[-1.5ex]{1.10}&$-10.35$&$-5.26$&$-2.46$\\[0.1cm] 
    Si$_{84}$(NH$_2)_{64}$&&$-18.51$&$-3.64$&$-0.69$ \\ 
    Si$_{84}$(OH)$_{64}$&\raisebox{1.5ex}[-1.5ex]{1.48}&$-18.52$&$-5.03$&$-2.43$\\[0.1cm]
    Si$_{165}$(NH$_2)_{100}$&&$-29.36$&$-3.62$&$-0.69$ \\ 
    Si$_{165}$(OH)$_{100}$&\raisebox{1.5ex}[-1.5ex]{1.85}&$-29.17$&$-5.10$&$-2.63$\\[0.1cm]
    Si$_{286}$(NH$_2)_{144}$&&$-42.41$&$-3.66$&$-0.82$ \\ 
    Si$_{286}$(OH)$_{144}$&\raisebox{1.5ex}[-1.5ex]{2.22}&$-43.06$&$-4.77$&$-2.72$\\[0.1cm]
    Si$_{455}$(NH$_2)_{196}$&&$-58.12$&$-3.58$&$-0.90$ \\ 
    Si$_{455}$(OH)$_{196}$&\raisebox{1.5ex}[-1.5ex]{2.59}&$-58.84$&$-4.67$&$-2.59$\\ \hline
    Si$_{10}$ in 1.5 ML Si$_3$N$_4$&&$-4.27$&$-4.02$&$-0.21$ \\ 
    Si$_{10}$ in 1.5 ML SiO$_2$&\raisebox{1.5ex}[-1.5ex]{0.72}&$-5.22$&$-5.67$&$-2.26$\\[0.1cm]
    Si$_{35}$ in 1.5 ML Si$_3$N$_4$&&$-9.23$&$-3.72$&$-0.73$ \\ 
    Si$_{35}$ in 1.5 ML SiO$_2$&\raisebox{1.5ex}[-1.5ex]{1.10}&$-10.86$&$-5.42$&$-2.76$\\[0.1cm] 
    Si$_{84}$ in 1.5 ML Si$_3$N$_4$&&$-16.60$&$-3.54$&$-0.93$ \\ 
    Si$_{84}$ in 1.5 ML SiO$_2$&\raisebox{1.5ex}[-1.5ex]{1.48}&$-19.30$&$-5.32$&$-2.89$\\[0.1cm]
    Si$_{165}$ in 1.5 ML Si$_3$N$_4$&&$-25.91$&$-3.54$&$-0.75$ \\ 
    Si$_{165}$ in 1.5 ML SiO$_2$&\raisebox{1.5ex}[-1.5ex]{1.85}&$-30.17$&$-5.19$&$-2.63$\\[0.1cm]
    Si$_{286}$ in 1.5 ML Si$_3$N$_4$&&$-37.24$&$-3.23$&$-1.02$ \\ 
    Si$_{286}$ in 1.5 ML SiO$_2$&\raisebox{1.5ex}[-1.5ex]{2.22}&$-43.89$&$-5.19$&$-2.74$\\ \hline   
  \end{tabular}
\end{table}
We would thus expect a nearly identical shift of the LNS i-Si electronic structure in accord with interface dipole theory. 

However, measurements of Si NCs \cite{Koe14} and nanowells (NWells) \cite{Koe18a,Koe19,Koe21}, and DFT calculations of Si NCs \cite{Koe08,Koe09a,Koe19} and nanowires (NWires) \cite{Koe18a} show a very different behavior. Energies of the lowest unoccupied MO (LUMO) $E_{\rm LUMO}$, and $E_{\rm HOMO}$ listed in Table \ref{tab1} show that Si NCs coated with SiO$_2$ experience a shift to higher $E_{\rm bind}$, and Si NCs coated in Si$_3$N$_4$ experience a shift to lower $E_{\rm bind}$. This contradiction to interface dipole theory can be resolved when looking at the quantum chemical properties of the chemical elements involved. Such elements are LNS i-Si as the cation providing electrons, and in particular N or O as the anion receiving such electrons. Relevant properties of the anions are the electronegativity (EN) and resulting ionicity of bond (IOB) to Si, the ionization energy $E_{\rm ion}$, the electron affinity for the neutral anion $X^0$, and the electron affinity of the anion ionized with one electron $X^-$, see Table \ref{tab2}. We focus on N and O as anions to explain the origin of the NESSIAS effect. 
\begin{table}[h!]
\caption{\label{tab2}Electronegativity (EN) of the ligand elements, resulting ionicity of bond (IOB) to Si, first ionization energy ($E_{\rm{ion}}$), electron affinity in neutral state ($X^0$) and ionized with one negative charge ($X^-$). Values are from \cite{HolWi95}, except $X^-$.}
\begin{tabular}{lrrrrr}
element&EN$^{\ast}$&IOB to Si&$E_{\rm{ion}}$&$X^0$&$X^-$\\
&&[\%]&[eV]&[eV]&[eV]\\ \hline
Si&1.74& 0 &8.15&$-1.38$& \\
B&2.01&2&8.30&$-0.28$&$+5.58$ \cite{Sche98}\,$^\dagger\,$\\
H&2.20&5&13.60&$-0.76$&$+5.85$ \cite{Guo90}\,$^\times$\\
C&2.50&13&11.26&$-1.26$&$+9.03$ \cite{Guo90}\,$^\times$\\
N&3.07&36&14.53&$+0.07$&$+8.30$ \cite{Huh83}\,$^\dagger\,$\\
O&3.50&54&13.36&$-1.46$&$+8.03$ \cite{Gins58}\,$^\times$\\
F&4.10&75&17.42&$-3.40$&$+7.69$ \cite{Guo90}\,$^\times$\\
S&2.44&12&10.36&$-2.08$&$+6.12$ \cite{Guo90}\,$^\times$\\
\end{tabular}
\vspace*{-0.25cm}
\begin{flushleft}
$^\ast$ Allred \& Rochow \ \ \ \ $^\dagger$ measured \ \ \ \ 
$^\times$ calculated
\end{flushleft}
\end{table}

\begin{figure}[h!]
\includegraphics[width=8.6cm,keepaspectratio]{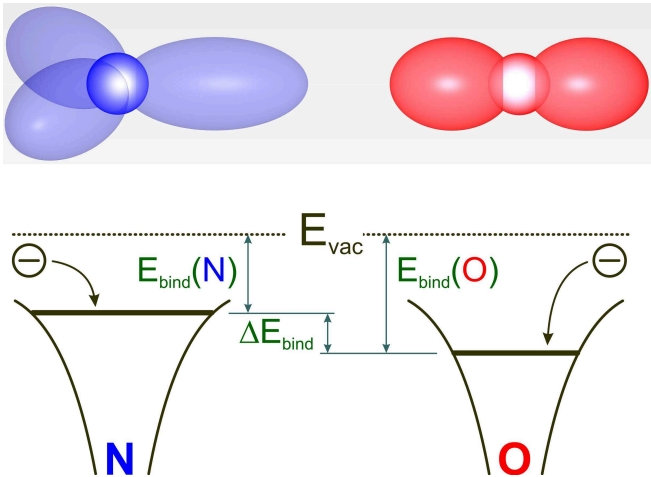}
\caption{\label{fig02}Relevant quantum chemical properties of N and O shown in spatial space and energy. The top graph shows the spatial extension of frontier occupied AOs. N and in particular O attract electrons from afar due to their high EN. The key difference between N and O is given by the \emph{local} electronic properties: N delocalizes extrinsic electrons due to its positive $X^0$ and $X^-$; O localizes such electrons with its more negative $X^0$ and $X^-$. The bottom graph shows the consequence in the energy picture: frontier AOs (and their anti-/non-bonding counterparts) shift to lower $E_{\rm bind}$ for Si$_3$N$_4$-embedding, while frontier AOs for SiO$_2$-embedding shift to higher  $E_{\rm bind}$.} 
\end{figure}

Table \ref{tab2} shows that N has a much more positive $X^0$ and still more positive $X^-$ than O, while its IOB to Si is ca. 2/3 of the value O provides. This lower IOB and thus the charge of the ICT \emph{per bond} $q_{\rm\,ICT}$ is nearly cancelled out by N in Si$_3$N$_4$ having 3/2 interface bonds to Si on average due to its trivalent configuration. Thus, values of the cumulative charge transferred to the main anions $\sum q_{\rm\, ICT}$ for NCs of same size embedded into 1 ML SiO$_2$- \emph{vs.} Si$_3$N$_4$ are nearly equal.
The situation changes when the dielectric embedding increases from 1 to 1.5 ML \cite{Koe19}, see bottom section of Table \ref{tab1}. For 1.5 ML SiO$_2$, $\sum q_{\rm\, ICT}$ increases for SiO$_2$-embedding, though this increase drops from ca. 16 \% for Si$_{10}$ NCs to ca. 2 \% for Si$_{286}$ NCs. We also see a slight shift of $E_{\rm{HOMO}}$ towards $E_{\rm vac}$ at Si$_{286}$ and Si$_{455}$ NCs,  and a shift of $E_{\rm{LUMO}}$ towards $E_{\rm vac}$ for Si$_{455}$ NCs, indicating an ICT saturation for 1 ML SiO$_2$, and thereby a saturation of NESSIAS at this NC size and embedding. Such saturation cannot be seen for an embedding in 1.5 ML SiO$_2$. Looking at NCs embedded in 1 ML Si$_3$N$_4$, we see that $\sum q_{\rm\, ICT}$ has lower values when compared to the same NC in 1 ML SiO$_2$, whereby the difference diminishes from ca. $-7$ \% for Si$_{10}$ NCs to ca. $-1$ \% for Si$_{455}$ NCs. There is no additional shift of $E_{\rm{HOMO}}$ and $E_{\rm{LUMO}}$ away from $E_{\rm vac}$, strongly suggesting that the NESSIAS effect is not saturated with 1 ML Si$_3$N$_4$ for Si$_{455}$ NCs. For 1.5 ML Si$_3$N$_4$-embedding, $\sum q_{\rm\, ICT}$ drops notably from a slight overshoot of ca. 2 \% for Si$_{10}$ NCs to $-12$ \% for a Si$_{286}$ NC, dropping accordingly more when compared to Si NCs in 1.5 ML SiO$_2$. Although $\sum q_{\rm\, ICT}$ is the lowest for 1.5 ML Si$_3$N$_4$-embedding, the values of $E_{\rm{HOMO}}$ and $E_{\rm{LUMO}}$ relative to $E_{\rm vac}$ remain virtually unchanged when compared to embedding in 1 ML Si$_3$N$_4$. 
We found the same result from DFT where two Si NCs  ranging from 0.7 to 1.9 nm size (Si$_{10}$ to Si$_{165}$) were calculated within one approximant, one of which was embedded in SiO$_2$, and the other in Si$_3$N$_4$, with 3 ML of dielectric between them \cite{Koe18a,Koe19}, see also Ref. \cite{SuppInfo} for details. We thus can state for SiO$_2$-embedding that the NESSIAS effect is rather spatially compact and comes out of saturation fairly quickly with increasing  LNS i-Si system size. For Si$_3$N$_4$-embedding, the NESSIAS effect is spatially distributed and the electronic shift is rather smooth, extending over a wider range of LNS i-Si system size. 
In order to understand this peculiar electronic structure, we look at a sketch combining spatial atomic orbital (AO) distribution and energy levels, see Fig. \ref{fig02}.
Arguably, the delocalizing impact of N \emph{vs.} the localizing impact of O onto their acquired electronic charge is the key to the NESSIAS effect. We can interpret a decreased $\sum q_{\rm\, ICT}$ for N as a partial reflection of $q_{\rm\, ICT}$ back into the LNS i-Si. These observations are supported by a decreasing $\sum q_{\rm\, ICT}$ per NC size when going from 1 to 1.5 ML Si$_3$N$_4$-embedding, being confirmed indirectly by experimental results of Si$_3$N$_4$- \emph{vs.} SiO$_2$-embedded Si NWells, see Section \ref{QC-SubEdges} and Fig. \ref{fig08}. 
The atomistic nature of the NESSIAS implies a short impact length in accord with other near-field effects such as significant electron tunneling \cite{Brar96}. For LNS i-Si, the extension of the NESSIAS is \emph{ca.} 1.3 to 1.8 nm per plane interface \cite{Koe18a,Koe19,Koe21}.

\subsection{\label{analytics}Analytic Relation of the NESSIAS Effect with the HOMO Energy}
We consider interfaces constituted by single bonds, such as between Si/SiO$_2$ and Si/Si$_3$N$_4$. A detailed derivation of the equations below and the use $E_{\rm ion}$, $X^0$ and $X^-$ of the interface main anion of the considered ligand group together with the average charge of the main anion $q_{\rm main}^{\rm avg}$ as boundary values are given in Appendix Section \ref{Append-Lambda-calc}. Here, we focus on results to express the binding energy of interface bonds as a function of the parameter $\Lambda_{\rm{main}}^{q({\rm main})}$. The average charge of the main anion constituting the ligands to Si NCs $q_{\rm main}^{\rm avg}$ is derived from DFT calculations and presents the only non-analytic input to $\Lambda_{\rm{main}}^{q({\rm main})}$. We calculate the parameter for a negative charge transfer to the main anion -- which is the most likely case -- 
\begin{eqnarray}\label{eqn-09}
\Lambda_{\rm{main}}^{q({\rm main})}&=&(1-|q_{\rm{main}}^{\rm{avg}}|)^{5/2} X_{\rm{main}}^0 +\\ &&\underbrace{|q_{\rm{main}}^{\rm{avg}}|^{5/2} X_{\rm{main}}^-}_{\rm{due\ to\ negative\ ionization}}\ \forall\ q_{\rm{main}}^{\rm{avg}}\leq 0\nonumber
\end{eqnarray}
and a positive charge transfer to the main anion of the ligand (or dielectric)
\begin{eqnarray}\label{eqn-10}
\Lambda_{\rm{main}}^{q({\rm main})}&=&(1-|q_{\rm{main}}^{\rm{avg}}|)^{5/2} X_{\rm{main}}^0 -\\ &&\underbrace{|q_{\rm{main}}^{\rm{avg}}|^{5/2} E_{\rm{ion,\,main}}}_{\rm{due\ to\ positive\ ionization}}\ \forall\ q_{\rm{main}}^{\rm{avg}}\geq 0\ .\nonumber
\end{eqnarray}
The parameter $\Lambda_{\rm{main}}^{q({\rm main})}$ describes the binding energy of the interface bond and is proportional to $E_{\rm HOMO}$, or the energy of the valence band maximum $E_{\rm V}$ for sufficiently large LNS i-Si systems like NWells evaluated in Section \ref{QC-SubEdges}. 
We can thus use $\Lambda_{\rm{main}}^{q({\rm main})}$ to predict the NESSIAS as a function of the embedding dielectric, thereby providing optimum combinations of Si and dielectrics per design to meet the desired functionality of VLSI electronic devices as briefly discussed below.
\begin{figure}[h!] 
\includegraphics[width=8.6cm,keepaspectratio]{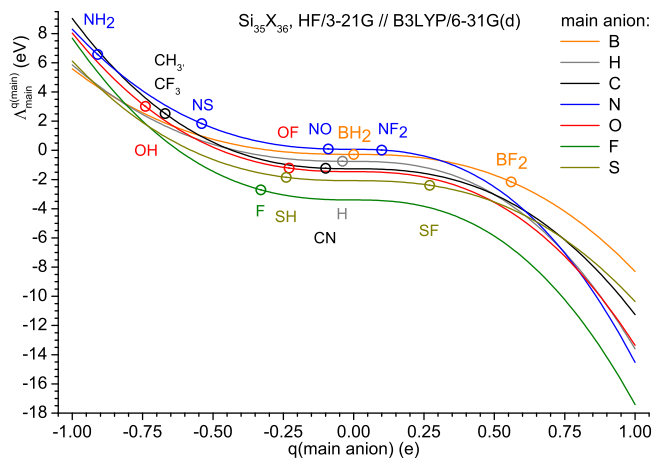}\hfill
\caption{\label{fig03}The parameter $\Lambda_{\rm main}^{q({\rm main})}$ shown as a function of transferred charge from LNS i-Si to main anion $q({\rm main})$ for all anionic terminations and their possible outer terminations, see Table \ref{tab2}. $\Lambda_{\rm main}^{q({\rm main})}$ for Si$_{35}$X$_{36}$ (X $=$ ligand) is shown per main anion for $q({\rm main})\in [-1;+1]$, with the respective data point per functional group termination derived from $q_{\rm main}({\rm DFT})$, see Fig. \ref{fig05} for corresponding data on electronic structure and $\Lambda_{\rm main}^{q({\rm main})}$. The difference in $\Lambda_{\rm main}^{q({\rm main})}$ as per main anion which describes the respective dielectric provides an estimate for the strength of the VB offset $\Delta E_{\rm V}$ between accordingly coated LNS i-Si sections, and can be used for VLSI device design.}  
\end{figure}
Fig. \ref{fig03} shows the result of Eqs. \ref{eqn-09} and \ref{eqn-10}, together with the data points of all ligands used in DFT calculations of Si$_{35}$ NCs (Section \ref{DFT}). 
Embedding in SiO$_2$ \emph{vs.} Si$_3$N$_4$ is indeed a good choice for a maximum NESSIAS. Coating LNS i-Si with Fluoride is even more attractive for a maximum electronic structure shift to higher $E_{\rm bind}$ as evident from the bigger difference in $\Lambda_{\rm{main}}^{q({\rm main})}$ to Si$_3$N$_4$-embedding. Such differences as per embedding dielectric of LNS i-Si are useful to predict the adequate combination of dielectrics to arrive at type II homojunctions required for VLSI field effect transistors (FETs), \emph{e.g.} using SiO$_2$- and Si$_3$N$_4$-embedding, or potentially at band-to-band tunneling (BTBT) FET devices by replacing SiO$_2$- with Fluoride-embedding.

Since Si and other semiconductors have a significant inter-atomic charge transfer, the NESSIAS effect will not be limited to a few atomic MLs, providing the basis for its use in VLSI electronics. Within the dielectric, a charge transfer over more than four Si--X (X$=$ O, N) MLs becomes unnotable in particular for SiO$_2$ embedding \cite{Koe18a,Koe19} because of the strong polar nature of the bonds and immobility of local charges due to charge localization and a rather wide bandgap of SiO$_2$ and Si$_3$N$_4$. 

\subsection{\label{DFT}Verification of the Analytic Relation with DFT Simulations}
\begin{figure}[h!] 
\includegraphics[width=4.25cm,keepaspectratio]{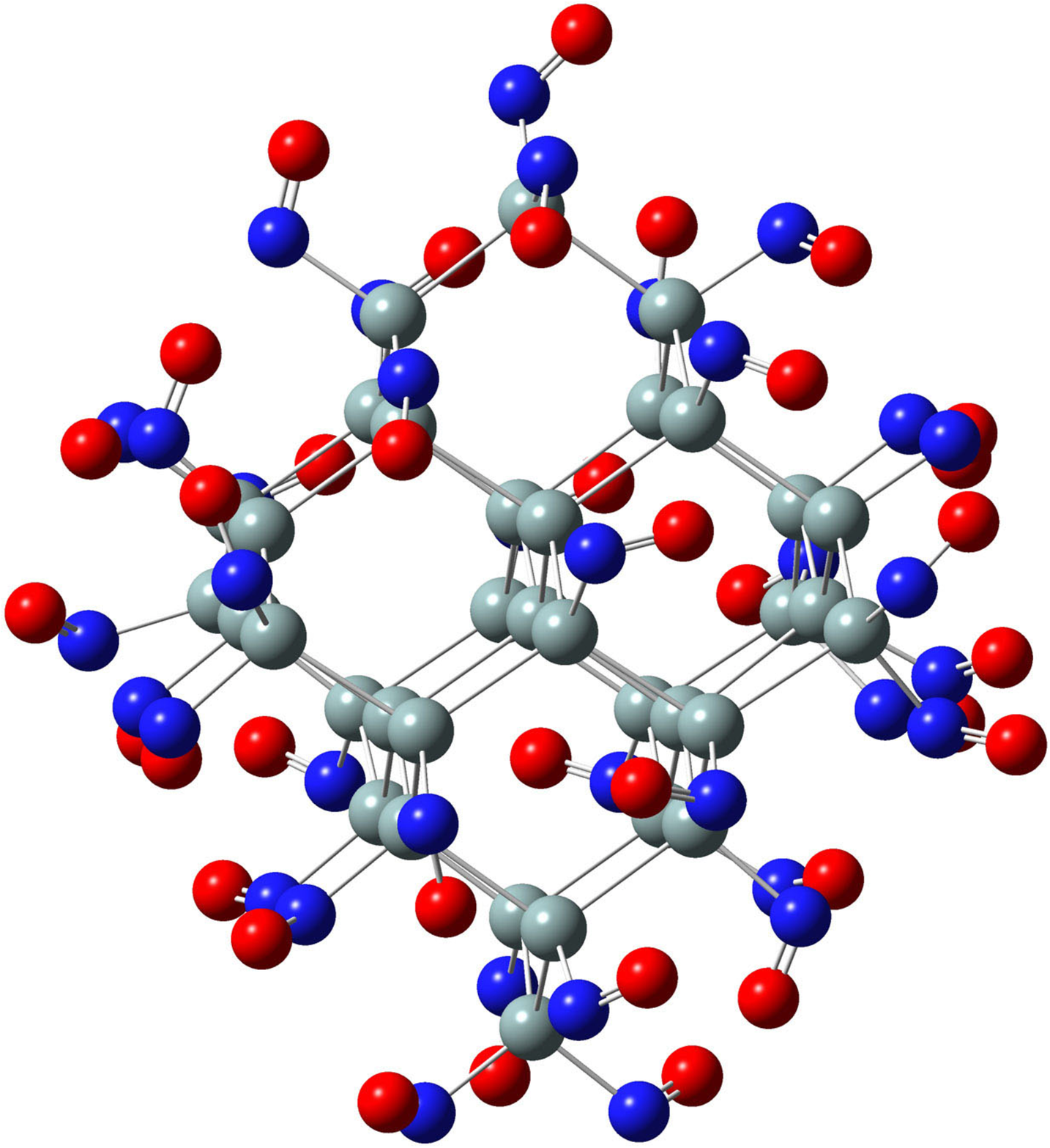}\hfill
\includegraphics[width=4.25cm,keepaspectratio]{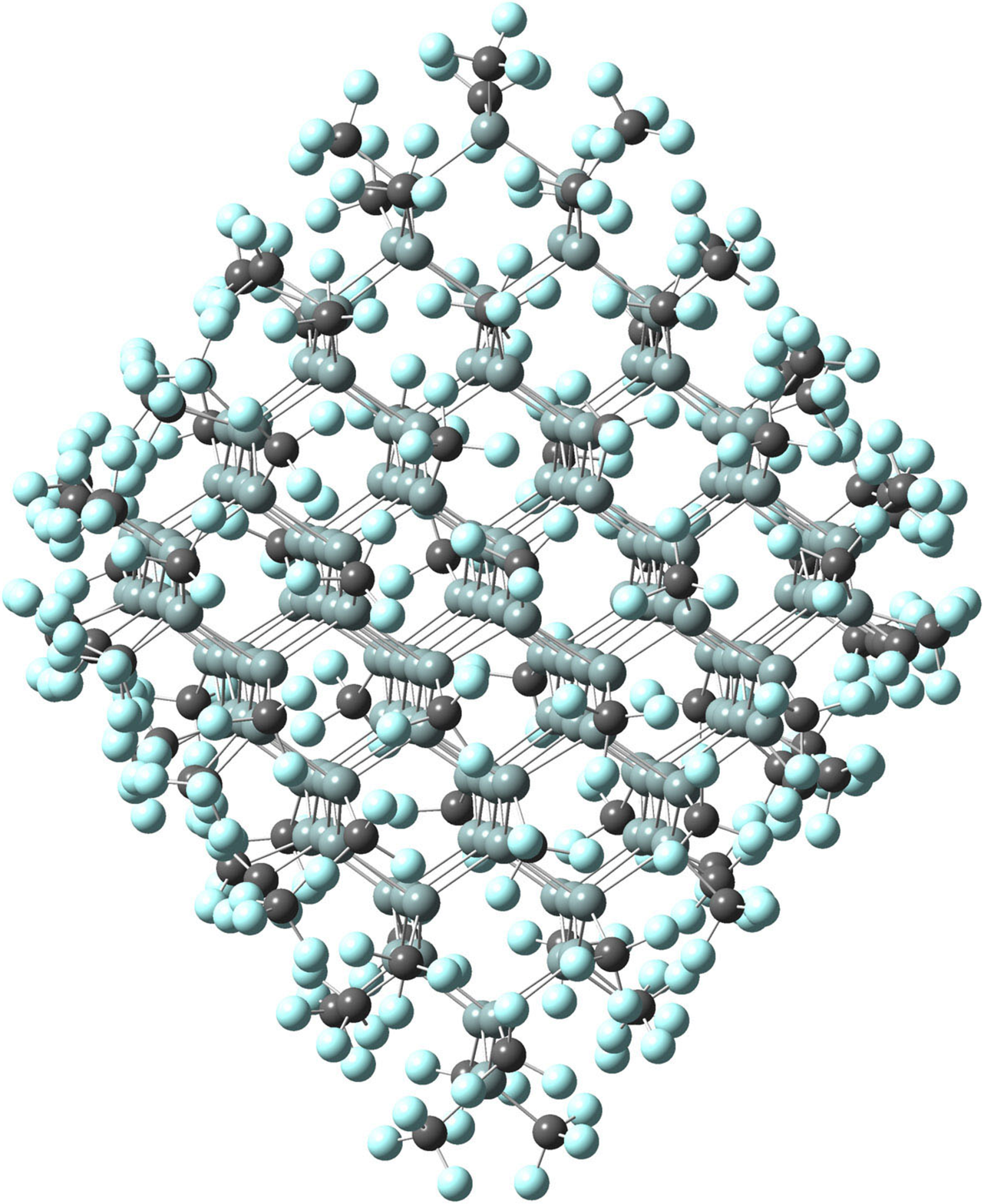}\\
{\bf (a)\hspace{3.77cm}(b)}\\[0.1cm]
\includegraphics[width=4.25cm,keepaspectratio]{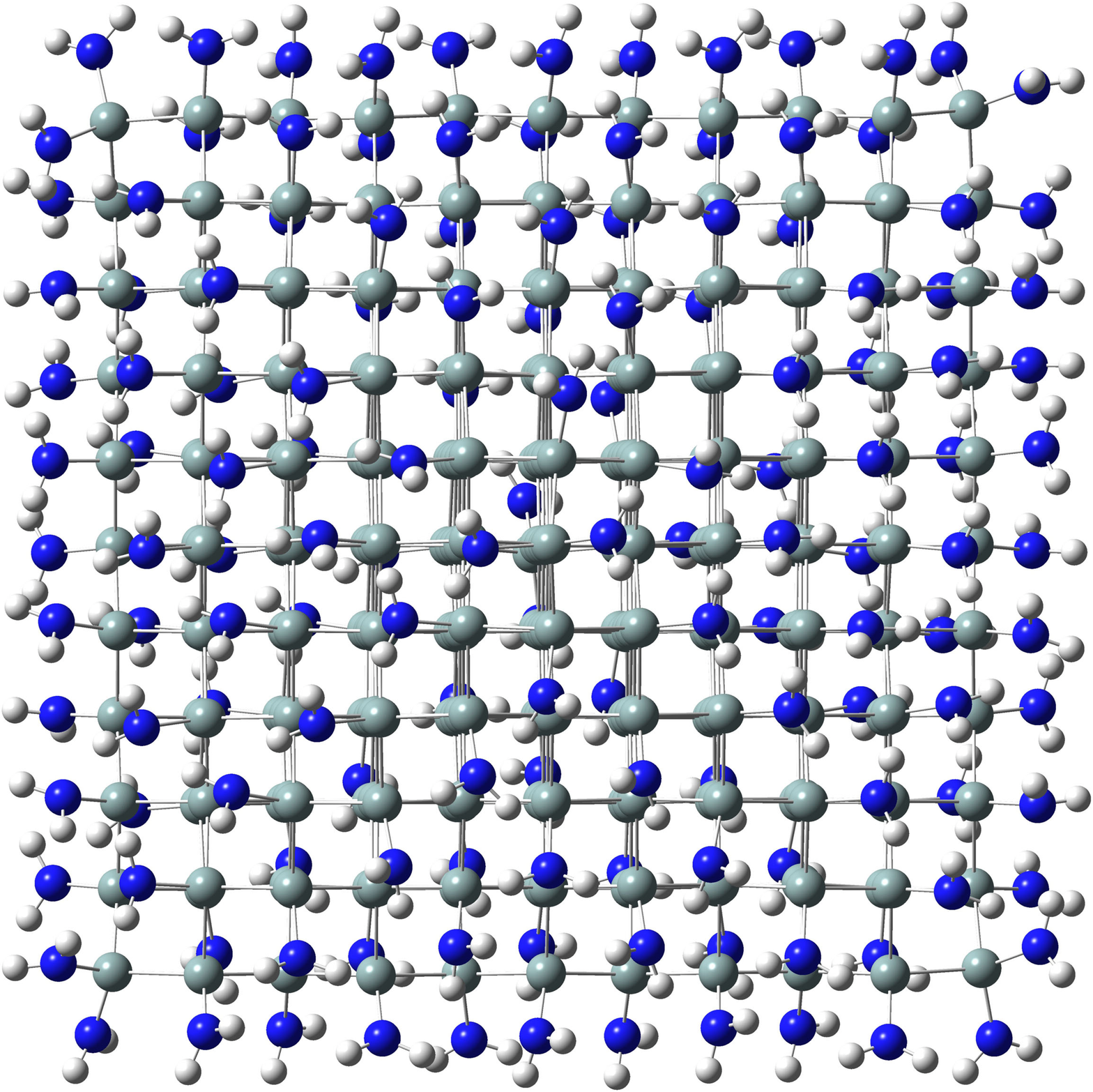}\hfill
\includegraphics[width=4.25cm,keepaspectratio]{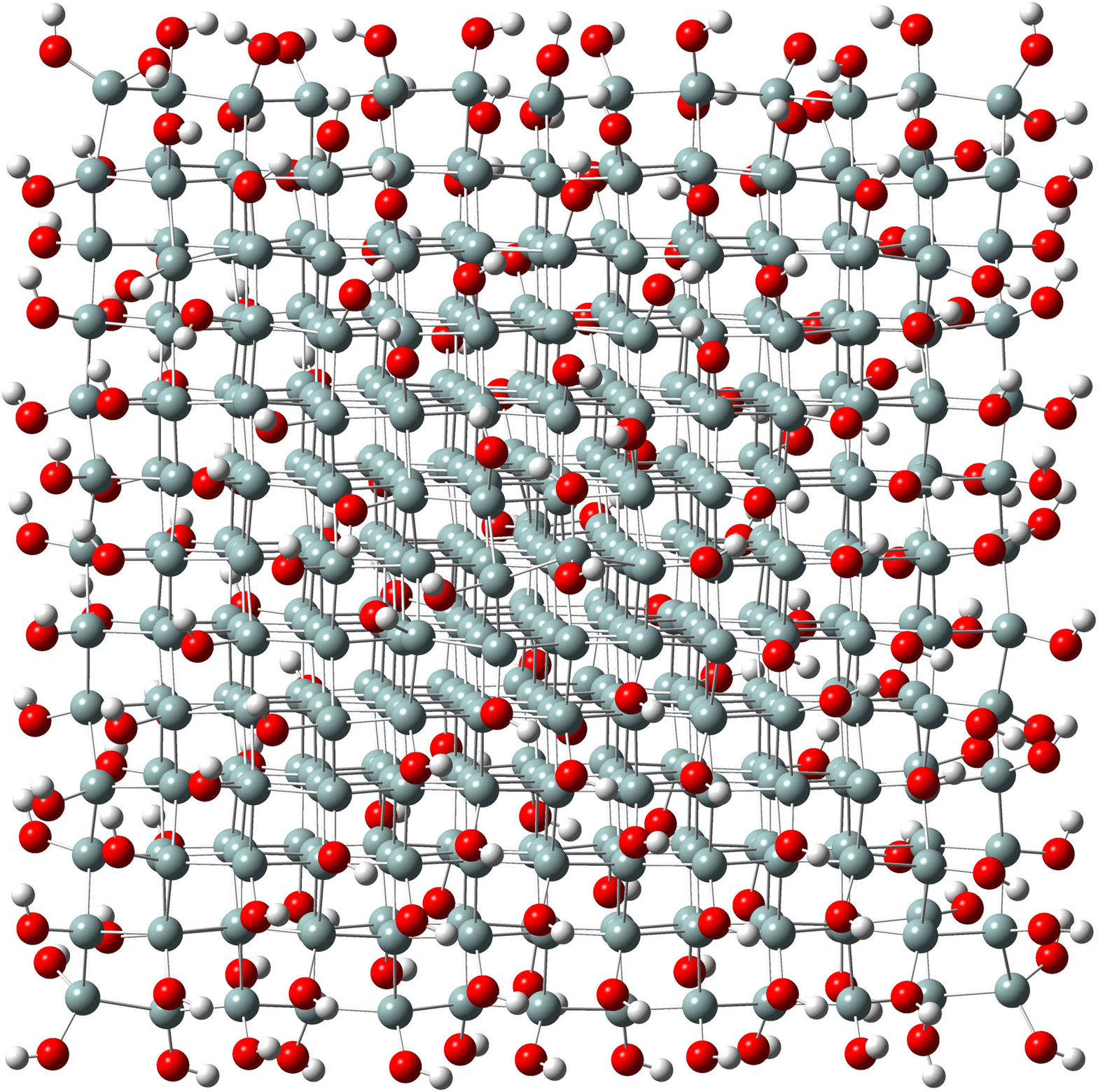}\\
{\bf (c)\hspace{3.93cm}(d)}
\caption{\label{fig04}
Examples of structurally optimized approximants of which DFT data are shown in Figs. \ref{fig03} and \ref{fig05} and in Ref. \cite{SuppInfo}: {\bf (a)} Si$_{35}$(NO)$_{36}$, {\bf (b)} Si$_{165}$(CF$_3$)$_{100}$, shown along the $\langle 110\rangle$ vector class, and {\bf (c)} Si$_{286}$(NH$_2$)$_{144}$, and {\bf (d)} Si$_{455}$(OH)$_{196}$, shown along the $\langle 001\rangle$ vector class. Atom colors are white (H), anthracite (C), blue (N), red (O), and gray (Si). Further  Data on other DFs and NC sizes can be found in Ref. \cite{SuppInfo}.}  
\end{figure}
DFT approximants consist of NCs fully terminated with one ligand type. The main anion forms the center of the ligand and the interface bond to the NC. Relevant properties of main anions are listed in Table \ref{tab2}. Fig. \ref{fig04} shows examples of NCs calculated by DFT. 
\begin{figure}[h!] 
\includegraphics[width=8.6cm,keepaspectratio]{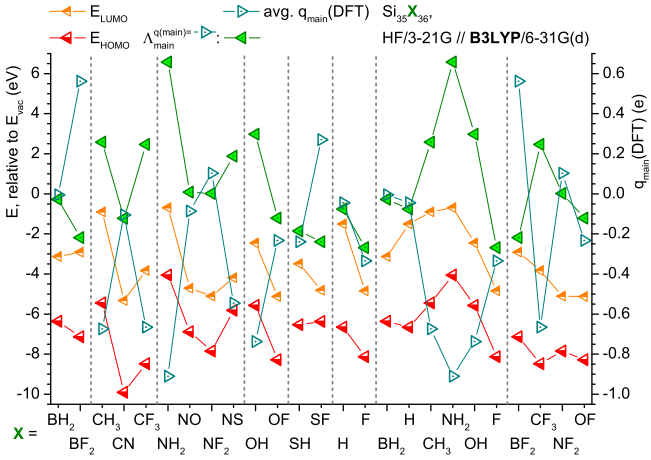}\hfill
\caption{\label{fig05}Energies $E_{\rm HOMO}$ and $E_{\rm LUMO}$ with $\Lambda_{\rm main}^{q({\rm main})}$ and $q({\rm main})$ (right scale) relative to $E_{\rm{vac}}$ for Si$_{35}$ NCs (1.1 nm size) as a function of complete surface termination noted at the abscissa. Electronic structure calculations were carried out as indicated in legend, see text for details. More data obtained by the CAM-B3LYP and HSE06 DFs for Si$_{35}$, and for Si NCs up to 2.6 nm size (Si$_{455}$) are available in Ref. \cite{SuppInfo}.}  
\end{figure}
The dependence $E_{\rm{HOMO}}\propto \Lambda_{\rm{main}}^{q({\rm main})}$ can be clearly seen for Si$_{35}$ NCs with a diameter of $d_{\rm NC}=1.1$ nm as a function of their surface termination in Fig. \ref{fig05}, where ligands are grouped in accord with their main anion, and specific outer terminations where applicable, arranged for increasing EN from left to right.
Recently, H-terminated 1.1 nm size Si NCs were processed by a top-down design \cite{Shir20}. 
For Si NCs fully terminated with BF$_2$, CF$_3$, NF$_2$, and OF ligands, $\Lambda_{\rm{main}}^{q({\rm main})}$ does not appear to follow $E_{\rm{HOMO}}$ as accurately. This behavior is explained in Ref. \cite{SuppInfo}.
All other terminations are described accurately by $\Lambda_{\rm{main}}^{q({\rm main})}\propto E_{\rm{HOMO}}$.

We extended calculations of Si$_{35}$ NCs to other hybrid DFs, the Heyd-Scuseria-Ernzerhof DF with its 2006 parametrization (HSE06) \cite{Kruk06}, and the B3LYP DF complemented with the Coulomb Attenuation Method (CAM-B3LYP) \cite{Yana04}. Detailed results of such calculations are listed in Ref. \cite{SuppInfo}, further corroborating the accuracy of $\Lambda_{\rm{main}}^{q({\rm main})}$ in predicting $E_{\rm HOMO}$.

\subsection{\label{SynchroExpr}Details of the NESSIAS Effect in Si NWells from Experiment}
\subsubsection{\label{QC-SubEdges}Calibrating the NESSIAS by Measuring the Intrinsic Valence Band Edge with Synchrotron UPS}
\begin{figure}[h!] 
\includegraphics[width=8.6cm,keepaspectratio]{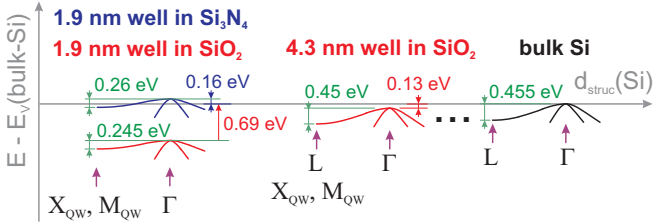}
\caption{\label{fig06}
Sketch of the electronic structure of the VB top with $\Delta E_{\rm V}^{\rm VanHove}$, evolving with decreasing $d_{\rm Well}$. Subbands are shown for bulk Si (black), for NWells in SiO$_2$ (red), and for NWells in Si$_3$N$_4$ (blue). Values of the VB shift relative to the VB edge of bulk Si were taken from  the respective least residual fit in Fig. \ref{fig08}. Values of $\Delta E_{\rm V}^{\rm VanHove}$ are shown in dark green, see to Fig. \ref{fig07}, and to Ref. \cite{SuppInfo} for more details.} 
\end{figure}

First data of SiO$_2$- and Si$_3$N$_4$-embedded Si NWells were obtained by measuring the VB maximum (leading edge) of bulk Si and embedded Si NWells using scan ensembles of synchrotron UPS \cite{Koe18a,Koe19,Koe21} with their statistic data, whereby the VB maximum is located at $\Gamma$ point; $E_{\rm V}=E_{\rm V}^{\Gamma}$. We now discuss the fine structure of UPS spectra, revealing a sub-edge which can be assigned to the Van Hove singularity of bulk Si at the L point in the BZ with its energy $E_{\rm V}^{\rm L}$ \cite{Chel89,Ley72}. For NWells with $d_{\rm Well}\leq 5$ nm, the VB subband structure becomes increasingly perturbed by QC. An assignment of subband Van Hove singularities near $E_{\rm V}$ becomes a function of $d_{\rm Well}$ and of specific high symmetry points in the electronic DOS along non-orthogonal $\mathbf{k}$-directions of the NWell plane, in particular X and M, see Ref. \cite{SuppInfo}. Fig. \ref{fig06} illustrates the change in VB electronic structure with the shrinking size of LNS i-Si using experimentally derived values. We focus on the energy offset between $\Gamma$ and L point for bulk Si, and between ${\Gamma}$ and X, M points for thin NWells, generally expressed by the term $\Delta E_{\rm V}^{\rm VanHove}$.

\begin{figure}[h!] 
\includegraphics[width=8.6cm,keepaspectratio]{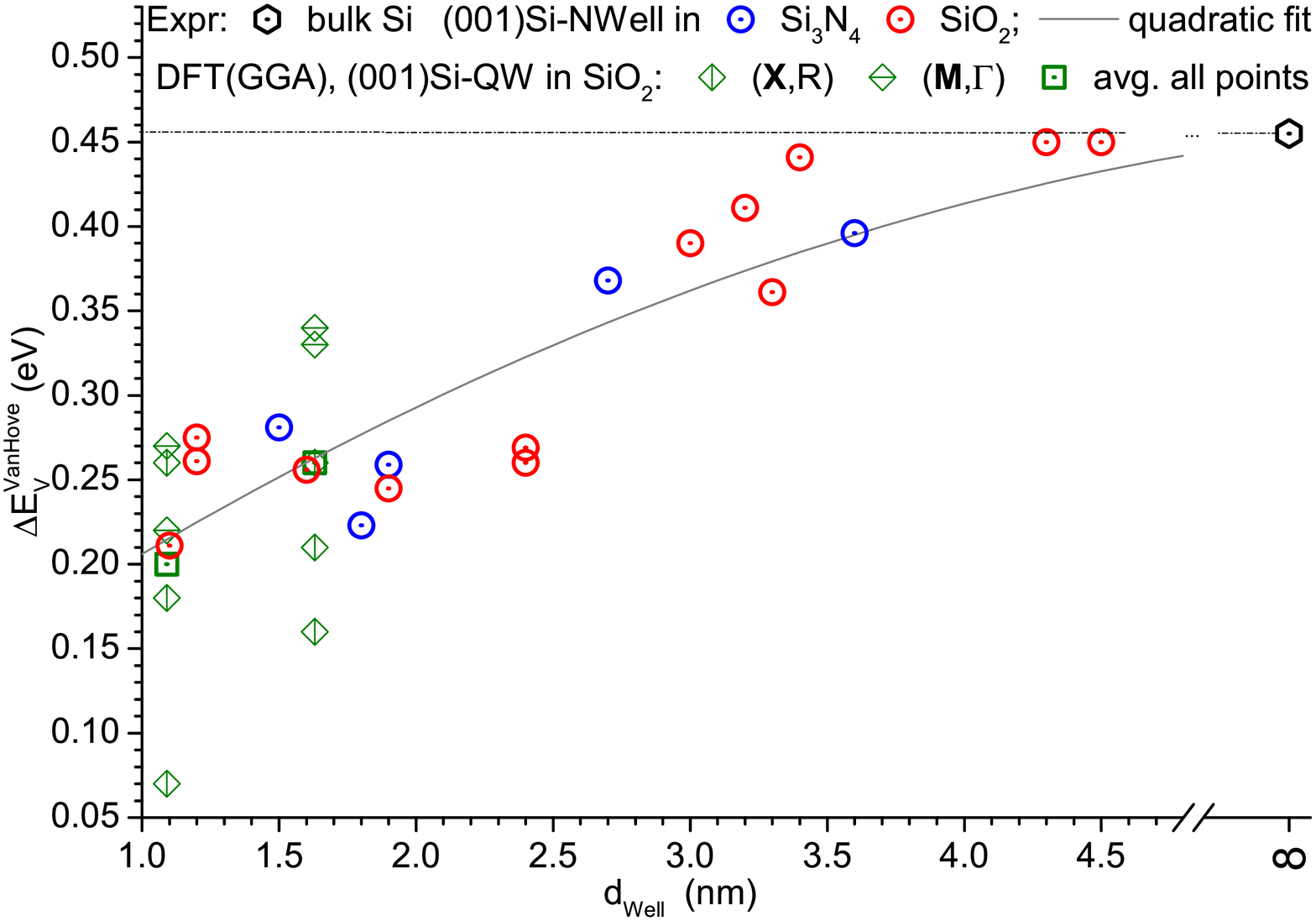}
\caption{\label{fig07}
Energy difference between VB-DOS extrema at $\Gamma$ point and adjacent Van Hove singularities $\Delta E_{\rm V}^{\rm VanHove}$ as function of $d_{\rm Well}$ embedded in SiO$_2$ (red dots) and in Si$_3$N$_4$ (blue dots). The gray line is a quadratic-hyperbolic least residuals fit, the black hexagonal symbol shows the bulk value of $\Delta E_{\rm V}^{\rm VanHove}$; please observe abscissa break. Green symbols denote Van Hove singularities for Si NWells in SiO$_2$ with $d_{\rm Well}=1.09$ and 1.63 nm, embedded in 1.92 nm thick SiO$_2$ barriers as calculated by DFT with periodic boundary conditions \cite{Carr02}. Small rhomboid symbols show local Van Hove singularities per $\mathbf{k}$-point, the big square symbol per $d_{\rm Well}$ shows their average value. For details on UPS measurement and data retrieval from published DFT calculations, see Ref. \cite{SuppInfo}.} 
\end{figure}

Fig. \ref{fig07} shows that $\Delta E_{\rm V}^{\rm VanHove}$ does not depend on the embedding dielectric as should be the case for a pure QC phenomenon with sufficiently high potential walls (i.e. band offsets between respective bulk phases) \cite{Schiff68}, \emph{cf.} Fig. \ref{fig10}(a). 
When QC sets in with decreasing $d_{\rm Well}$, the energy offset between VB-DOS extrema at $\Gamma$ and adjacent Van Hove singularities $\Delta E_{\rm V}^{\rm VanHove}$ diminishes as $E_{\rm V}^{\Gamma}$ is shifted to higher binding energies. Since the adjacent Van Hove singularities already have higher binding energies with respect to the VB maximum, they do not experience a notable energy shift due to QC even for the minimum energetic distance of $\Delta E_{\rm V}^{\rm VanHove}\approx 0.2$ eV for NWells with $d_{\rm Well}=1.1$ nm, see Fig. \ref{fig07}. We can therefore take $\Delta E_{\rm V}^{\rm VanHove}$ as a good estimate for the shift of $E_{\rm V}^{\Gamma}=E_{\rm V}$ to higher binding energies due to QC only. 
Interestingly, DFT-GGA calculations of Si NWell superlattices (SLs) with 1.9 nm thick SiO$_2$ barriers \cite{Carr02} yield very similar values of $\Delta E_{\rm V}^{\rm VanHove}(d_{\rm Well})$, \emph{cf.} Fig. \ref{fig07} and discussion in Ref. \cite{SuppInfo}. 
Yet, the energy offset $\Delta E_{\rm V}^{\rm VanHove}$ derived from UPS does not reflect the unperturbed QC case due to its finite energy difference, accounting for some uncertainty of the Van Hove singularity with higher binding energy \cite{Schiff68} as the reference level for $\Delta E_{\rm V}^{\rm VanHove}$. Still, this deviation on the order of 1 meV is further diminished by the DOS in the Brillouin zone (BZ) at L point in bulk Si being \emph{ca.} five-fold bigger as compared to the DOS at $\Gamma$ point, see to UPS data for surface-cleaned bulk Si (Fig. S7, Ref. \cite{SuppInfo}). Local deviations due to stress/strain and the deviation of $d_{\rm Well}$ of ca. $\pm 0.2$ nm around its nominal value have a more significant impact on $\Delta E_{\rm V}^{\rm VanHove}$, see Fig. \ref{fig07}. Hence, $\Delta E_{\rm V}^{\rm VanHove}$ as measured by UPS is an accurate estimate of VB QC. We introduce a VB maximum as a function of NWell thickness,
\begin{eqnarray}
E_{\rm V}^{\rm QC}(d_{\rm Well})&=&E_{\rm V}({\rm bulk\ Si})
\,+\,\Big[E_{\rm V}^{\Gamma}({\rm bulk\ Si})\\
&&-E_{\rm V}^{\rm L}({\rm bulk\ Si})\Big]-\Delta E_{\rm V}^{\rm VanHove}\nonumber\\
&=&5.17\,{\rm eV}\,+\,0.455\,{\rm eV}-\Delta E_{\rm V}^{\rm VanHove}\, .\nonumber
\end{eqnarray} 
In Fig. \ref{fig07}, a quadratic-hyperbolic least residuals fit was used in accord with QC theory \cite{Schiff68} to estimate the shift of $E_{\rm V}$ as a function of $d_{\rm Well}$. 

\begin{figure}[h!] 
\includegraphics[width=8.6cm,keepaspectratio]{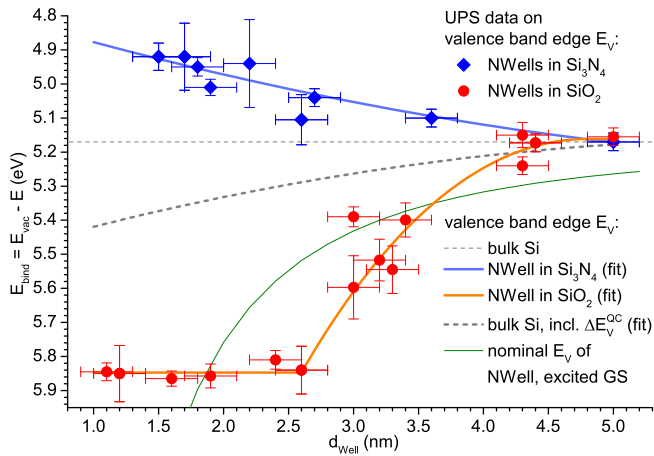}
\caption{\label{fig08}
Synchrotron UPS data of the VB maxima of Si NWells embedded in SiO$_2$ (red symbols) and Si$_3$N$_4$ (blue symbols). Error bars show standard deviations in thickness and energy. Orange and light blue lines provide a least square residual fit to $E_{\rm V}(d_{\rm Well})$ for SiO$_2$- and Si$_3$N$_4$-embedding, respectively. The thin dashed gray line shows $E_{\rm V}$ of bulk Si, the thick dashed gray line shows $E_{\rm V}$ with intrinsic QC $E_{\rm V}^{\rm QC}(d_{\rm Well})$, see text  for details. The green line shows $E_{\rm V}$ of the nanowell for photon absorption (including exciton binding energy, excluding lattice relaxation) as per existing theory.}  
\end{figure}
We present all UPS data together with  $E_{\rm V}^{\rm QC}(d_{\rm Well})$ in Fig. \ref{fig08}. The energy offset due to the NESSIAS as per embedding dielectric is clearly visible, saturating for SiO$_2$-embedding for $d_{\rm Well}\leq 2.6$ nm, while steadily growing with decreasing $d_{\rm Well}$ for Si$_3$N$_4$-embedding. The VB offset between 1.6 nm thick Si NWells coated with Si$_3$N$_4$ \emph{vs.} SiO$_2$ is $\Delta E_{\rm V}\approx 0.95$ eV, facilitating charge carrier separation on a massive scale in analogy to a steep p/n junction induced by impurity doping. Before discussing the full electronic structure of embedded Si NWells with experimental data, we revisit the discussion of the quantitative NESSIAS impact due to Si$_3$N$_4$ \emph{vs.} SiO$_2$.

\begin{figure}[h!] 
\includegraphics[width=8.6cm,keepaspectratio]{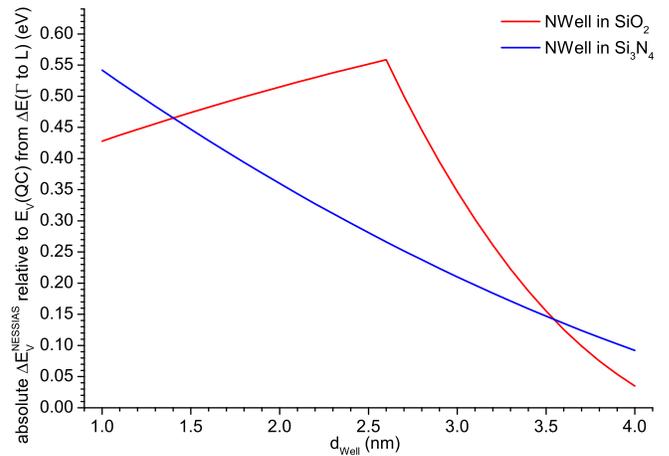}
\caption{\label{fig09}
Absolute values of $\Delta E_{\rm V}^{\rm NESSIAS}$ as a function of $d_{\rm Well}$ relative to $E_{\rm V}^{\rm QC}(d_{\rm Well})$, using the least square residual fits of VB edges of Fig. \ref{fig08}. The NESSIAS becomes saturated for NWells in SiO$_2$ around $d_{\rm Well}=2.6$ nm, but not for $d_{\rm Well}\geq 1.0$ nm for NWells in Si$_3$N$_4$, see text for further discussion.}  
\end{figure}

With $E_{\rm V}^{\rm QC}(d_{\rm Well})$, we have a true reference level to investigate on a quantitative base how the NESSIAS affects Si NWells per embedding dielectric. To that effect, we calculate the absolute value of the difference between $E_{\rm V}^{\rm QC}(d_{\rm Well})$ and the VB maximum $E_{\rm V}(d_{\rm Well})$ per embedding dielectric, \emph{viz.} $|\Delta E_{\rm V}^{\rm NESSIAS}|=|E_{\rm V}^{\rm QC}(d_{\rm Well})-E_{\rm V}(d_{\rm Well})|$ \emph{cf.} Fig. \ref{fig09}.
As already emerging from Fig. \ref{fig08}, $|\Delta E_{\rm V}^{\rm NESSIAS}|$ becomes saturated for $d_{\rm Well}\leq 2.6$ nm for NWells in SiO$_2$. The strong localization of extrinsic electrons from the NWell at the O atoms in SiO$_2$ has two effects. Due to the small localization volume which is limited to the immediate proximity of O atoms, see Section \ref{qualiExplain}, these atoms undergo electrostatic screening. Thereby, the ICT from the NWell to SiO$_2$ is self-limiting. The NWell undergoes a rather strong positive ionization, increasing the attractive Coulomb force which works against the ICT. Together with the screened O atoms, the cumulative ICT charge $\sum q_{\rm\, ICT}$ thus \emph{decreases} for $d_{\rm Well}< 2.6$ nm, whereby its partition per Si NWell atom still increases. The latter statement is straightforward to see when comparing $|\Delta E_{\rm V}^{\rm NESSIAS}|$ of $d_{\rm Well}= 2.6$ nm with its value at half the NWell thickness. There, $|\Delta E_{\rm V}^{\rm NESSIAS}|$ decreased from its maximum of 0.56 eV at $d_{\rm Well}= 2.6$ nm not by about 0.28 eV as would be the case for a constant NWell ionization, but by a mere 0.10 eV, accounting for a further increase of the positive ionization per Si NWell atom. We note, though, that the electronic DOS of the VB over energy is not constant. Further decrease in $|\Delta E_{\rm V}^{\rm NESSIAS}|$ may occur due to an increased DOS at the energy where $E_{\rm V}^{\rm QC}(d_{\rm Well})$ resides. 
The situation is very different for Si$_3$N$_4$-embedding, where $|\Delta E_{\rm V}^{\rm NESSIAS}|$ is constantly increasing for a decreasing $d_{\rm Well}$, getting close to the saturation limit of SiO$_2$-embedding. This behavior is in accord with the quantum-chemical properties of N, see Section \ref{qualiExplain}. It is uncertain whether we can proceed to Si$_3$N$_4$-embedded NWells with $d_{\rm Well}< 1.1$ nm to find a saturation limit in experiment should it exist. What we do know from experiment is that the constant increase of $|\Delta E_{\rm V}^{\rm NESSIAS}|$ with decreasing $d_{\rm Well}$ results from the electron-attractive, yet delocalizing nature of N which does not suffer from a saturation due to a much bigger volume which can be utilized by N to accommodate extrinsic electrons. As discussed in Section \ref{qualiExplain}, this peculiar behavior also results in a part of the ICT to be reflected back into the NWell as a consequence of the decreased binding energy of extrinsic electrons at N atoms, thereby shifting the electronic structure of the NWell and LNS i-Si in general towards $E_{\rm vac}$.

We have stated in Section \ref{qualiExplain} that 1 ML SiO$_2$ can keep the NESSIAS saturated for a ratio of LNS i-Si atoms per interface bond of $N_{\rm IF}/N_{\rm Si}=196/455\approx 0.43$ for Si NCs. Fig. \ref{fig08} shows that this saturation is left around $d_{\rm Well}=2.6$ nm which corresponds to ca. 19 atomic ML of (001)Si or 38 Si atoms and eight interface bonds per unit cell area, resulting in $N_{\rm IF}/N_{\rm Si}=8/38\approx 0.21$. Assuming an exponential decay in charge transfer through SiO$_2$ when moving away from the interface, we arrive at the estimate of ca. 3 ML or 0.9 nm SiO$_2$ \cite{Koe14,Koe18a} after which the NESSIAS does not increase significantly anymore by increasing the thickness of SiO$_2$. While this is good news for the VLSI technology where ultrathin SiO$_2$ layers are required to chemically passivate LNS i-Si and to establish the primary layer of gate dielectrics, there is little chance to extend a saturated NESSIAS in Si beyond $d_{\rm Well}=2.6$ nm. For Si$_3$N$_4$-embedding, the delocalizing impact of N yields to a NESSIAS smeared out over $d_{\rm Well}$, with no saturation present for $d_{\rm Well}\geq 1.5$ nm.

\subsubsection{\label{QC-FullPic}A First Glimpse on the Full NESSIAS Picture} 
\begin{figure}[t!] 
\includegraphics[width=8.6cm,keepaspectratio]{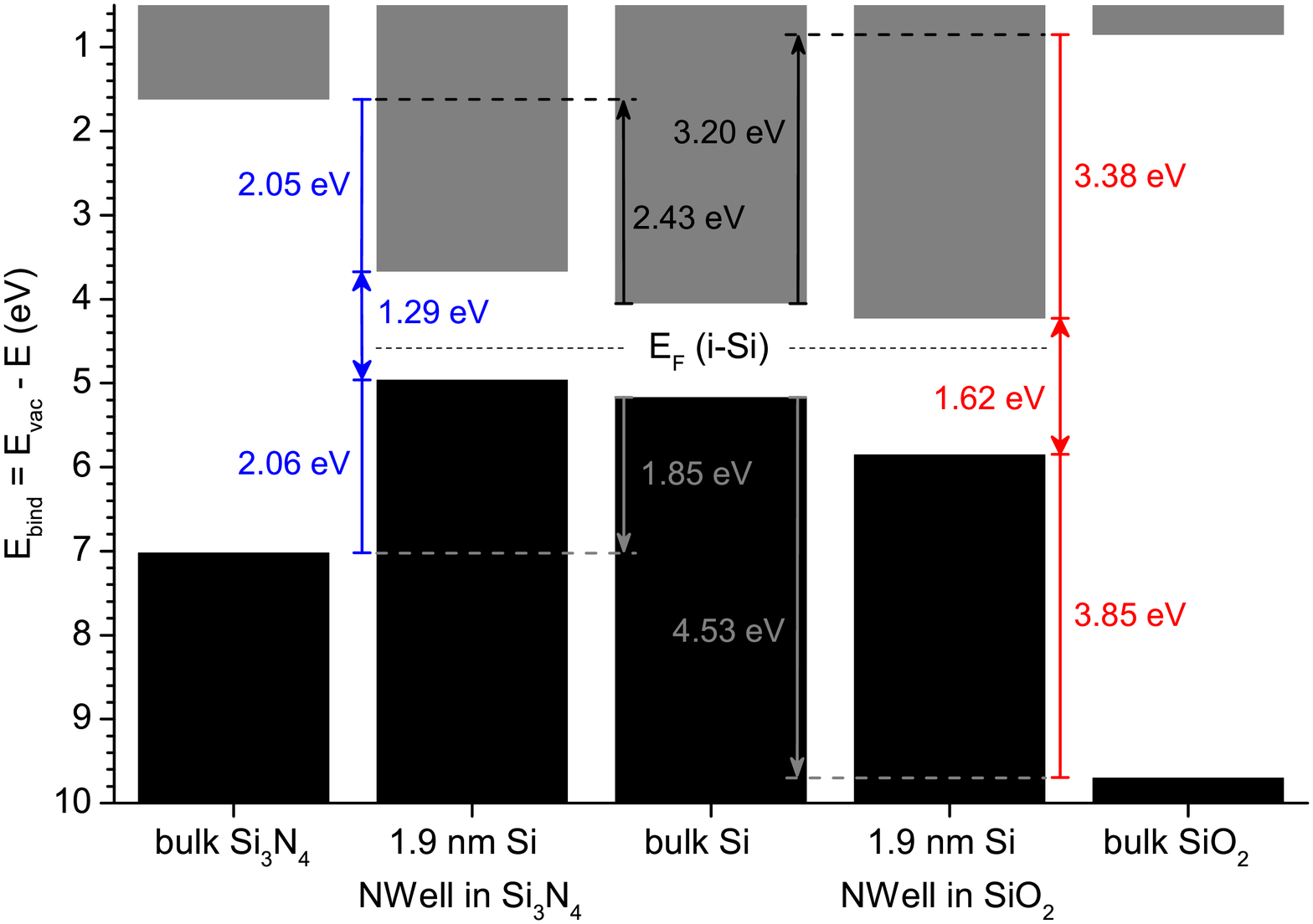}\\[-0.1cm]
{\bf (a)}\\[0.1cm]
\includegraphics[width=8.6cm,keepaspectratio]{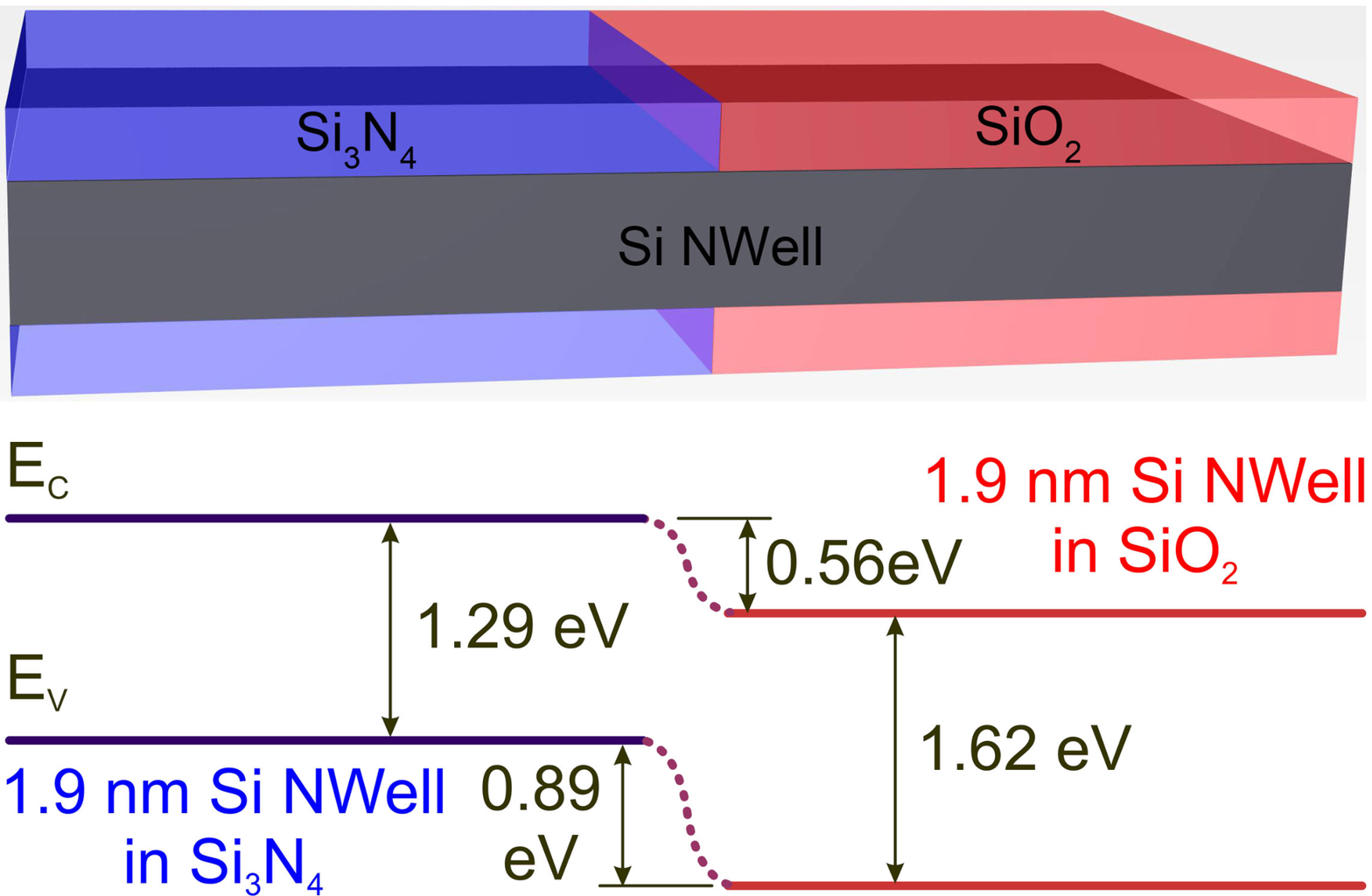}\\
{\bf (b)}\\
\caption{\label{fig10}
(a) Regions of fundamental band gaps for -- left to right -- bulk Si$_3$N$_4$ \cite{Kura81,Iqba87,Keis99,Koe18a}, a 1.9 nm Si-NWell in Si$_3$N$_4$, intrinsic bulk Si \cite{Nic82,Koe18a}, a 1.9 nm Si-NWell in SiO$_2$, and bulk SiO$_2$ \cite{Ibach74,Nic82,Keis99,Koe18a}. Band offsets to dielectrics are shown by labelled red (blue) lines for NWells in SiO$_2$ (Si$_3$N$_4$), and in gray/black for bulk Si. Band edges given contain values of bulk solids from experiment, data for NWells were obtained by UPS and XAS-TFY, see to text, Ref. \cite{SuppInfo} and to \cite{Koe18a,Koe19,Koe21} with respective supplements. (b) Layer system comprising a 1.9 nm thick Si NWell coated with 1.0 nm Si$_3$N$_4$ and SiO$_2$ (top), and its band diagram with energies in units of eV (bottom), derived from experimental data in graph (a).}  
\end{figure}
We recently started to measure the energy of the conduction band edge $E_{\rm C}$ of Si NWells embedded in SiO$_2$ by means of X-ray Absorption Spectroscopy in Total Fluorescence Yield mode (XAS-TFY) \cite{Koe21}. An initial sample of a Si$_3$N$_4$-coated NWell gave us a very first glimpse on the electronic structure of Si NWells of comparable thickness as a function of their embedding. We thus focus here on two NWell samples with $d_{\rm Well}=1.9\pm 0.2$ nm and the respective embedding. Details of other SiO$_2$-coated NWells can be found in \cite{Koe21} and its supporting information, representative scans and statistical data of both samples used here can be found in Ref. \cite{SuppInfo}. The obtained experimental data for $E_{\rm V}$ and $E_{\rm C}$ are shown together with the bulk phases of Si, SiO$_2$, and Si$_3$N$_4$ in Fig. \ref{fig10}(a). Fig. \ref{fig10}(b) shows a combination of both NWell samples, with the resulting band diagram as per experimental data from Fig. \ref{fig10}(a). Obviously, we can induce a strong type II homojunction \cite{Harr05} in LNS i-Si merely by coating with SiO$_2$ \emph{vs.} Si$_3$N$_4$.

The band structures of bulk solids shown in Fig. \ref{fig10}(a) yield to further insights into the NESSIAS effect. The band offsets between the bulk phases of Si and SiO$_2$ as well as Si and Si$_3$N$_4$ are somewhat asymmetric. A strong shift of $E_{\rm C}$, $E_{\rm V}$ to higher binding energies occurs for SiO$_2$. For Si$_3$N$_4$, we see a smaller shift to lower binding energies. Both observations are in accord with our discussion of O and N \emph{vs.} Si in Section \ref{qualiExplain}. We thus can track the NESSIAS effect all the way to the bulk phase of the respective dielectric. As a consequence, the entire electronic structure of LNS i-Si such as a sufficiently thin NWell moves towards a more symmetric band offset with the respective dielectric when the NESSIAS effect becomes significant. 

From above discussion, the question of a LNS i-Si size limit arises below which it starts to behave like the embedding dielectric to a degree where they depart from the properties of semiconducting Si. 
Further research efforts will be required to provide a thickness estimate around which the NESSIAS effect can be put to best use for technological applications from the viewpoint of LNS i-Si system size. Moreover, recent DFT calculations \cite{Koe21} showed that other group IV semiconductors such as C and Ge -- and presumably their alloys SiC and SiGe -- strongly respond to the NESSIAS, offering a much broader impact on VLSI device design beyond LNS Si. 

\section{\label{WrapUp} Conclusions}
We quantitatively demonstrated that an energy shift of electronic states defining the CB and VB edge of low nanoscale intrinsic Si (LNS i-Si) exists, being considerably different from commonly assumed energy positions. The underlying Nanoscale Electronic Structure Shift Induced by Anions at Surfaces (NESSIAS) is brought about by the embedding dielectric, whereby N in the form of Si$_3$N$_4$ and O in the form of SiO$_2$ are of particular scientific and technological interest. We explained the underlying quantum chemical processes of the NESSIAS which mainly reside with the anion of the dielectric, namely its ability to attract extrinsic electrons such as from LNS i-Si from afar (electronegativity EN defining the ionicity of bond to Si), and its ability to localize extrinsic electrons in its immediate vicinity (electron affinity $X$). Both, O and N attract extrinsic electrons from i-Si via an interface charge transfer. While such electrons get strongly localized at O with its highly negative $X^0$, they get delocalized around N due to its $X^0$ being positive. Therefore, electronic states experience a shift to higher binding energies with O as an interface anion, while the delocalization of extrinsic electrons in the immediate vicinity of N reflects such states partly back to the LNS i-Si, lowering their binding energy. Due to these quantum chemical properties, the NESSIAS  saturates rather quickly as a function of SiO$_2$ thickness, while still increasing for Si$_3$N$_4$-coatings beyond three atomic MLs.
The NESSIAS thus results in an electronic structure shift towards (away from) $E_{\rm vac}$ for Si$_3$N$_4$- (SiO$_2$-) embedding. Hence, LNS i-Si systems like NWires \cite{Koe21} or NWells can be flooded by holes (Si$_3$N$_4$-coated) or electrons (SiO$_2$-coated), a property which readily offers the formation of a type II (p/n) homojunction in LNS i-Si. 

We introduced an analytic parameter as a function of the ionization of the main interface anion $\Lambda_{\rm{main}}^{q({\rm main})}$ which correctly predicts the HOMO energy $E_{\rm HOMO}$ of the LNS i-Si system as a function of the main interface anion, and verified its accuracy with three different hybrid DFs. As main anions to Si, we evaluated the entire 1$^{\rm st}$ period of the periodic table and S, with all of these main anions terminated by all possible outer atoms to form ligands. 
Using $\Lambda_{\rm{main}}^{q({\rm main})}$ to predict the NESSIAS as a function of the dielectric coating, we obtain optimum combinations of Si and dielectrics per design to meet the desired functionality of VLSI electronic devices. 
Such predictions are useful in particular for ultra-low power VLSI devices and ultimate cryogenic electronics as with NESSIAS, no dopants are required for a p/n junction to exist.

Band edges of 1.9 nm thick Si NWells measured by UPS and XAS-TFY revealed band offsets due to Si$_3$N$_4$- \emph{vs.} SiO$_2$-coating of $\Delta E_{\rm C}=0.56$ eV and $\Delta E_{\rm V}=0.89$ eV. The band gap energies were $E_{\rm gap}=1.29$ eV for Si$_3$N$_4$- and $E_{\rm gap}=1.62$ eV for SiO$_2$-embedding, respectively.  Looking at the evolution of experimental $E_{\rm C}$ and $E_{\rm V}$ values with decreasing $d_{\rm Well}$, we showed that the band offsets of the NWells to the respective dielectric get more symmetric. It emerged that the absolute position of $E_{\rm C}$ and $E_{\rm V}$ residing with the respective dielectric is also a function of the quantum chemical properties at the origin of the NESSIAS. Detecting the NESSIAS in DFT calculations and in experiment requires the evaluation of $E_{\rm C}$ and $E_{\rm V}$ on an \emph{absolute} energy scale. \emph{e.g.} their position to $E_{\rm vac}$. The impact length for a saturation of the NESSIAS is ca. $1.5\pm 0.2$ nm per plane interface, resulting in $\leq 3.0\pm 0.4$ nm thick Si NWells and $\leq 5.5\pm 0.8$ nm thick Si NWires \cite{Koe21}. This short impact length and the lack of published DFT calculations and measurements with an \emph{absolute} energy calibration are the likely reasons why the NESSIAS has not been detected until recently \cite{Koe08,Koe14,Koe18a}. With VLSI approaching ultrathin fins and NWires, device dimensions are currently advancing into the thickness ranges mentioned above. It should therefore be of high interest to the VLSI research community to carry out device-related research, in particular in the light of the prospective ultra-low power demand and low temperature functionality given by the NESSIAS effect.

\begin{acknowledgments}
D.~K. acknowledges the 2018 Theodore-von-K{\`a}rm{\`a}n Fellowship of RWTH Aachen University, Germany. M.~F., N.~W. and J.~K.  acknowledge support by the Impulse and Networking Fund of the Helmholtz Association. D.H. acknowledges funding via a Heisenberg grant by the Deutsche Forschungsgemeinschaft (DFG,
German Research Foundation), project \#434030435. I.P., F.B., and E.M. acknowledge funding from EUROFEL project (RoadMap Esfri) and thank Federico Salvador (IOM-CNR) for technical support. The authors acknowledge Elettra Sincrotrone Trieste for providing access to its synchrotron radiation facilities (through proposals \#20165089, \#20180054, \#20190057, \#20200079, \#20205084, \#20215072, and \#20215081), and thank L. Sancin for technical support at Elettra, and Simone Dal Zilio for advise and use of the Micro and Nano fabrication facility FNF at IOM-CNR. D.~K. and S.~C.~S. acknowledge the National Computational Infrastructure (NCI) for the generous computing resources allocated on the Gadi supercomputer, Australian National University, Canberra, Australia.\\
\end{acknowledgments} 

\begin{appendix}
\section{\label{Append-Lambda-calc}Derivation of the quantum Chemical Parameter $\Lambda$ to describe $\mathbf E_{\rm\mathbf{HOMO}}$ for Various Interface Ligands/Dielectrics}
We set out by relating $E_{\rm ion}$, $X^0$ and $X^-$ of the interface main anion of the ligand group to the average charge transferred from the LNS Si to the main anion $q_{\rm main}^{\rm avg}$:
\begin{eqnarray}\label{eqn-01A}
\Lambda_{\rm{main}}^{q({\rm main})}&=&(1-|q_{\rm{main}}^{\rm{avg}}|)X_{\rm{main}}^0 +\\ &&\underbrace{|q_{\rm{main}}^{\rm{avg}}|X_{\rm{main}}^-}_{\rm{due\ to\ negative\ ionization}}\ \forall\ q_{\rm{main}}^{\rm{avg}}\leq 0\nonumber
\end{eqnarray}
\begin{eqnarray}\label{eqn-02A}
\Lambda_{\rm{main}}^{q({\rm main})}&=&(1-|q_{\rm{main}}^{\rm{avg}}|)X_{\rm{main}}^0 -\\ &&\underbrace{|q_{\rm{main}}^{\rm{avg}}|E_{\rm{ion,\,main}}}_{\rm{due\ to\ positive\ ionization}}\ \forall\ q_{\rm{main}}^{\rm{avg}}\geq 0\nonumber
\end{eqnarray}
For $q_{\rm{main}}^{\rm{avg}}=0$, we arrive at the trivial result that the regular electron affinity $X^0$ applies. The term $E_{\rm{ion,\,main}}$ stands for the ionization energy of the main anion which is counted positive towards $E_{\rm{vac}}$; $X^0$ and $X^-$ are counted negative (positive) when going below (above) $E_{\rm{vac}}$. For avoiding confusion with the signs of $q_{\rm{main}}^{\rm{avg}}$, we use its absolute value. For $q_{\rm{main}}^{\rm{avg}}\geq 0$, we include $E_{\rm{ion,\,main}}$ due to $X_{\rm{main}}^0$ being increasingly substituted by $E_{\rm{ion,\,main}}$, since a localized electron would have to fill up the vacant valency before it starts to present a real negative charge in the sense of an electron affinity. The influence of the cation is implicitly included in above equations via $q_{\rm{main}}^{\rm{avg}}$ which depends on $E_{\rm ion}$ of the nominal cation (semiconductor structure), and in a few cases from its electron affinities $X^0,\,X^-$ when charge is transferred from the embedding material to the LNS semiconductor system, as is the case for SF$_2$ or BF$_2$ ligands in Fig. \ref{fig05}. This atypical behavior is due to the peculiar quantum chemical nature of F in ligands, see Ref. \cite{SuppInfo} for further discussion.

\begin{figure}[h!] 
\includegraphics[width=8.6cm,keepaspectratio]{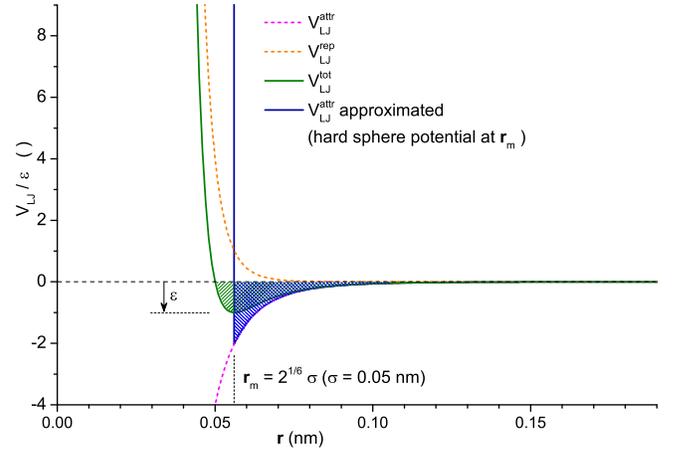}
\caption{\label{fig11}
Lennard-Jones potential $V_{\rm LJ}$ and hard-sphere potential to approximate the attractive part $V_{\rm LJ}^{\rm attr}$ used for deriving $\Lambda_{\rm{main}}^{q({\rm main})}$. The normalized $V_{\rm LJ}/\epsilon$ (dark green line) consists of a repulsive nuclear term $V_{\rm LJ}^{\rm rep}$ (orange dashed line) and an attractive Coulomb term $V_{\rm LJ}^{\rm attr}$ (magenta dashed line). The hard-sphere potential has $V\rightarrow\infty$ for ${\mathbf r}\leq {\mathbf r}_m$, and $V_{\rm LJ}^{\rm attr}$ for ${\mathbf r}\geq {\mathbf r}_m$ (blue line). The green hatched area marks the potential well of $V_{\rm LJ}$, the blue hatched area marks the well area of the approximated $V_{\rm LJ}^{\rm attr}$ used in integration of $q_{\rm main}^{\rm avg}$ over $\mathbf r$.} 
\end{figure}

Equations \ref{eqn-01A} and \ref{eqn-02A} establish the principal relation to describe the NESSIAS effect. We now work out the dependence of $\Lambda_{\rm{main}}^{q({\rm main})}$ on $q_{\rm{main}}^{\rm{avg}}$. 
To this end, we need a Coulomb term which is a function of $q_{\rm main}^{\rm avg}$ localized at the main anion, leading to a Coulomb functional (Coulomb field) of a quasi-point charge distributed in a local atomic potential of the main anion, \emph{viz.} $\mathbf{F_{\rm\mathbf Coul}}(q_{\rm main}^{\rm avg}[\mathbf{r}])$. This approach requires a local potential to accomodate $q_{\rm main}^{\rm avg}$ which is given by the Lennard-Jones potential \cite{Elli98} 
\begin{eqnarray}\label{eqn-03A}
V_{\rm LJ}({\mathbf r})&=&
4\epsilon\left[\left(\frac{\sigma}{\mathbf r}\right)^{12} -\left(\frac{\sigma}{\mathbf r}\right)^6\right], 
\end{eqnarray}
with $\epsilon$ being the minimum energy of the potential well, and $\sigma$ being the inter-particle distance which refers to the inter-electron distance of the electron pair forming the single bond per main anion to the Si NC, see Fig. \ref{fig11}. The potential minimum $V_{\rm LJ}/\epsilon=-1$ occurs at ${\mathbf r}_m= 2^{1/6}\sigma$, with $\sigma$ set to 0.05 nm as an example.
In our investigation of the anion impact, we will focus on the \emph{attractive} part of $V_{\rm LJ}^{\rm attr}$, namely  $V_{\rm LJ}^{\rm attr}=-(\sigma/{\mathbf r})^6$ which can enable or prevent the localization of an external electron, depending on the charge state and outer termination of the main anion. Since we do not have the exact inter-particle distance of the two electrons forming the Si-anion bond (nor does this make much sense from a quantum mechanical viewpoint), and $\epsilon$ varies as per anion type and its charge state, we work with proportionalities in analogy to solving a partial fraction decomposition by comparison of coefficients \cite{OxMathUsrGde04}. The scaling of such proportionality relations are given by $E_{\rm ion}$, $X^0$, and $X^-$ at $q_{\rm main}^{\rm avg}=+1$, 0, and $-1$, respectively. We approximate $V_{\rm LJ}^{\rm attr}({\mathbf r})$ by a hard-sphere potential of $V\rightarrow\infty$ at ${\mathbf r}={\mathbf r}_m = 2^{1/6}\sigma$, and $V({\mathbf r})=V_{\rm LJ}^{\rm attr}({\mathbf r})$ for ${\mathbf r}\geq 2^{1/6}\sigma$. 
The exact and the approximated well potentials integrated over their outward spatial extension $\mathbf r$ should be possibly identical to ensure an adequate description of charge localization. 
As shown in Appendix Section \ref{WellAreaCompare}, the spatial extension of the well of the approximated $V_{\rm LJ}^{\rm attr}$ deviates from $V_{\rm LJ}$ by a mere 2.9 \%.
No boundary conditions are needed as integration limits since we scale the proportionality relations with quantum chemical properties as per above, merely requiring the correct description of all variables in powers of ${\mathbf r}$. For the accumulated (or depleted) charge $q_{\rm main}^{\rm avg}[\mathbf{r}]$, we integrate over its location, \emph{viz.}  
\begin{eqnarray}\label{eqn-04A}
q_{\rm main}^{\rm avg}[\mathbf{r}]&=&q_{\rm main}^{\rm avg}\int V_{\rm LJ}^{\rm attr} d{\mathbf r}\\
&\propto&\frac{q_{\rm main}^{\rm avg}}{{\mathbf r^5}}\nonumber
\end{eqnarray}

Due to their comparatively big distance as compared to the localized anionic charge, we can approximate the Coulomb field of the main anion onto NC atoms as a point charge, yielding
\begin{eqnarray}\label{eqn-05A}
{\mathbf F_{\rm\mathbf Coul}}(q_{\rm main}^{\rm avg}[\mathbf{r}])&=&e\,\frac{q_{\rm main}^{\rm avg}}{4\pi\varepsilon_0\,{\mathbf r^2}}\,\propto\,\frac{q_{\rm main}^{\rm avg}}{{\mathbf r^2}}, 
\end{eqnarray}
where $e$ is the elementary charge presenting the probe charge into which the Coulomb field exerts its impact, and $\varepsilon_0$ is the permittivity constant of vacuum. We thus get ${\mathbf F_{\rm\mathbf Coul}}(q_{\rm main}^{\rm avg}[\mathbf{r}])$ describing an \emph{energy} gradient 
which will become more apparent below.

Next, we provide a gauge for the interface impact on an atomic scale, \emph{i.e.} an area density of potential energy due to charge accumulation/depletion at the main anion which influences the electronic structure of the NC atoms. Such density is given by the quotient of the Coulomb impact onto the NC and the charge localization at the main anion:
\begin{eqnarray}\label{eqn-06A}
\frac{{\rm field\ due\ to\ }q_{\rm main}^{\rm avg}{\rm\ at\ NC}}{{\rm location\ of\ }q_{\rm main}^{\rm avg}{\rm\ at\ anion}}&=&\frac{{\mathbf F_{\rm\mathbf Coul}}(q_{\rm main}^{\rm avg})|_{\rm NC}}{q_{\rm main}^{\rm avg}[\mathbf{r}]|_{\rm anion}}\nonumber\\
&\propto&\frac{\displaystyle 1/{\displaystyle\mathbf r^2}}{\displaystyle 1/{\displaystyle\mathbf r^5}}\,=\,{\mathbf r^{5/2}}
\end{eqnarray}
Equation \ref{eqn-06A} presents a Coulomb pressure onto the NC 
in analogy to a force constant 
conveying the exertion of force onto a mechanical system: 
An increased $q_{\rm main}^{\rm avg}$ results in an increased ${\mathbf F_{\rm\mathbf Coul}}(q_{\rm main}^{\rm avg})$ which will impact upon the probe charge $e$, whereby the location of $q_{\rm main}^{\rm avg}$ resides with $V_{\rm LJ}^{\rm attr}$, thereby scaling the impact of the quantum chemical properties of the main anion, \emph{viz.} $X_{\rm{main}}^0,\,X_{\rm{main}}^-$, and $E_{\rm ion}$.
With relation \ref{eqn-06A} describing an area density of Coulomb energy, we substitute $q_{\rm main}^{\rm avg}$ in Eqs. \ref{eqn-01A} and \ref{eqn-02A} with $(q_{\rm main}^{\rm avg})^{5/2}$, ending up with
\begin{eqnarray}\label{eqn-07A}
\Lambda_{\rm{main}}^{q({\rm main})}&=&(1-|q_{\rm{main}}^{\rm{avg}}|)^{5/2} X_{\rm{main}}^0 +\\ &&\underbrace{|q_{\rm{main}}^{\rm{avg}}|^{5/2} X_{\rm{main}}^-}_{\rm{due\ to\ negative\ ionization}}\ \forall\ q_{\rm{main}}^{\rm{avg}}\leq 0\nonumber
\end{eqnarray}
\begin{eqnarray}\label{eqn-08A}
\Lambda_{\rm{main}}^{q({\rm main})}&=&(1-|q_{\rm{main}}^{\rm{avg}}|)^{5/2} X_{\rm{main}}^0 -\\ &&\underbrace{|q_{\rm{main}}^{\rm{avg}}|^{5/2} E_{\rm{ion,\,main}}}_{\rm{due\ to\ positive\ ionization}}\ \forall\ q_{\rm{main}}^{\rm{avg}}\geq 0\, ,\nonumber
\end{eqnarray}
which are equivalent to Eqs. \ref{eqn-09} and \ref{eqn-10} in the main article.\\

\subsection{\label{WellAreaCompare}Comparison of Well Potentials -- $\mathbf V_{\rm LJ}$ vs. Approximated $\mathbf V_{\rm LJ}^{\rm attr}$}
For comparing both well areas, we integrate the respective potential over $\mathbf r$ with its integration limits, yielding
\begin{eqnarray}\label{eqn-A01}
A_{\rm Well}(V_{\rm LJ})&=&\int\limits_{\sigma}^{\infty}\left(\frac{\sigma}{{\mathbf r}}\right)^{12}-\left(\frac{\sigma}{{\mathbf r}}\right)^6\,d{\mathbf r}\\
&=& \left[-\frac{\sigma^{12}}{11\cdot{\mathbf r}^{11}}+\frac{\sigma^6}{5\cdot{\mathbf r}^5}\right]_{\sigma}^{\infty}\nonumber\\
&=&-\frac{6}{55}\sigma\nonumber
\end{eqnarray}
for the well formed by the Lennard-Jones potential, and 
\begin{eqnarray}\label{eqn-A02}
A_{\rm Well}(V_{\rm LJ}^{\rm attr}\,{\rm approx.}) &=&\int\limits_{2^{1\!/6}\sigma}^{\infty}-\left(\frac{\sigma}{{\mathbf r}}\right)^6\,d{\mathbf r}\\
&=& \left[\frac{\sigma^6}{5{\mathbf r}^5}\right]_{2^{1\!/6}\sigma}^{\infty} \nonumber\\
&=&-\frac{1}{5\cdot 2^{\,5/6}}\sigma\nonumber
\end{eqnarray}
for the well described by the hard sphere potential approximating $V_{\rm LJ}^{\rm attr}$. With $A_{\rm Well}(V_{\rm LJ})$ being the exact potential, we arrive at 
\begin{eqnarray}\label{eqn-A03}
\frac{A_{\rm Well}(V_{\rm LJ}^{\rm attr}\,{\rm approx.})}{A_{\rm Well}(V_{\rm LJ})}&=&
\frac{-55\sigma}{-6\cdot 5\cdot\sigma\cdot 2^{\,5/6}}, 
\end{eqnarray}
yielding 1.0289, \emph{i.e.} a deviation of the approximated well area from the exact Lennard-Jones well of ca. 2.9 \%.

\end{appendix}

\bibliography{RevTeX_Manuscript_v01}

\end{document}